\documentclass[a4paper,11pt,oneside,openany,enumerate]{memoir}
\usepackage{datetime}
\settrimmedsize{297mm}{210mm}{*}
\settypeblocksize{634pt}{448.13pt}{*} 
\setulmargins{4cm}{*}{*} 
\setlrmargins{*}{*}{1.5} 
\setmarginnotes{17pt}{51pt}{\onelineskip} 
\setheadfoot{\onelineskip}{2\onelineskip} 
\setheaderspaces{*}{2\onelineskip}{*} 
\checkandfixthelayout

\usepackage{fouriernc}
\usepackage[T1]{fontenc}
\usepackage{lipsum}

\OnehalfSpacing 
\setsecnumdepth{subsection} 
\maxsecnumdepth{subsubsection}
\usepackage{calc,soul,fourier} 
\makeatletter 
\newlength\dlf@normtxtw 
\newsavebox{\feline@chapter} 
\newcommand\feline@chapter@marker[1][4cm]{%
    \sbox\feline@chapter{%
        \resizebox{!}{#1}{\fboxsep=1pt%
            \colorbox{gray}{\color{white}\thechapter}%
        }}%
        \rotatebox{90}{%
                \resizebox{%
                \heightof{\usebox{\feline@chapter}}+\depthof{\usebox{\feline@chapter}}}%
            {!}{\scshape\so\@chapapp}}\quad%
        \raisebox{\depthof{\usebox{\feline@chapter}}}{\usebox{\feline@chapter}}%
} 
\newcommand\feline@chm[1][4cm]{%
    \sbox\feline@chapter{\feline@chapter@marker[#1]}%
    \makebox[0pt][c]{
        \makebox[1cm][r]{\usebox\feline@chapter}%
    }}
\makechapterstyle{daleifmodif}{

    \renewcommand\printchapternum{\null\hfill\feline@chm[2.5cm]\par}

} 
\makeatother 
\chapterstyle{daleifmodif}

\makepagestyle{myvf} 
\makeoddfoot{myvf}{}{\thepage}{} 
\makeevenfoot{myvf}{}{\thepage}{} 
\makeheadrule{myvf}{\textwidth}{\normalrulethickness} 
\makeevenhead{myvf}{\small\textsc{\leftmark}}{}{} 
\makeoddhead{myvf}{}{}{\small\textsc{\rightmark}}
\makeindex

\usepackage{import}
\usepackage{amsfonts}                
\usepackage[centertags]{amsmath}                            
\usepackage{amssymb}                
\usepackage{amsthm}                     
\usepackage{graphicx}
\usepackage[utf8]{inputenc}                                
\usepackage[T1]{fontenc}                                   
\usepackage[english]{babel}

\usepackage{color}
\usepackage{xcolor}
\usepackage{listings}
 
\usepackage{caption}
\DeclareCaptionFont{white}{\color{white}}
\DeclareCaptionFormat{listing}{\colorbox{gray}{\parbox{16cm}{#3}}}
\usepackage{hyperref}                           
\usepackage{microtype}

\usepackage{lettrine}

\newtheoremstyle{plain}
  {\topsep}   
  {\topsep}   
  {\nopagebreak\itshape}  
  {0pt}       
  {\nopagebreak\bfseries} 
  {.}         
  {5pt plus 1pt minus 1pt} 
  {} 

\newtheoremstyle{definition}
  {0.5\topsep}   
  {0.5\topsep}   
  {\nopagebreak\normalfont}  
  {0pt}       
  {\nopagebreak\bfseries} 
  {.}         
  {5pt plus 1pt minus 1pt} 
  {}

\newtheoremstyle{kommentar}
  {}   
  {}   
  {}  
  {}       
  {\bfseries} 
  {:}         
  {5pt plus 1pt minus 1pt} 
  {}

\theoremstyle{definition}

\theoremstyle{kommentar}

\usepackage{amsmath,amssymb}
\usepackage{textgreek}
\usepackage{hyperref}
\usepackage{subcaption}
\usepackage{geometry}
\usepackage{booktabs} 
\usepackage{multirow}
\usepackage{titlecaps}
\usepackage{pdfpages}
\usepackage{eucal}
\newcommand{\citeA}[1]{\@citeX{[Ref.~}{#1}{]}}
\newcommand{\citeB}[1]{\@citeX{Reference~}{#1}{}}
\usepackage[sorting=none,style=numeric-comp,mincitenames=2,maxbibnames=2,maxcitenames=2,uniquelist=false]{biblatex}
\usepackage{filecontents,notoccite}
\AtEveryCitekey{\clearfield{url}}
\DeclareSourcemap{
 \maps[datatype=bibtex ]{
   \map{
     \step [ fieldset = url , null ]
     \step [ fieldset = doi , null ]
     \step [ fieldset = issn, null ]
     \step [ fieldset = isbn, null ]
     }
  }
}
\usepackage{booktabs}
\usepackage{graphicx}
\usepackage{gensymb}
\usepackage{url}
\usepackage{imakeidx}
\makeindex[intoc=true,columns=1]

\usepackage{enumitem}
\usepackage{enumitem}
\usepackage{libertine}
\hypersetup{
    colorlinks=true,
    linkcolor=black,
    filecolor=black,      
    urlcolor=black,
    citecolor=black,
    pdftitle={Spin Dynamics in Patterned Magnetic Multilayers with Perpendicular Magnetic Anisotropy},
    pdfpagemode=FullScreen,
    }

\usepackage[utf8]{inputenc}
\usepackage{textcomp}
\inputencoding{utf8}
\DeclareUnicodeCharacter{00A0}{ }
\DeclareUnicodeCharacter{0229}{~}
\usepackage{authblk}
\usepackage{filecontents}
\DeclareCiteCommand{\fullcite}
  {\usebibmacro{prenote}}
  {\clearfield{url}%
   \clearfield{pagetotal}%
   \clearfield{edition}%
  \clearfield{number}%
   \clearfield{labelyear}%
   \usedriver
     {\DeclareNameAlias{sortname}{default}}
     {\thefield{entrytype}}}
  {\multicitedelim}
  {\usebibmacro{postnote}}
  
\usepackage[colorinlistoftodos]{todonotes}

\title{Spin Dynamics in Patterned Magnetic Multilayers with Perpendicular Magnetic Anisotropy}

\author[1]{Mateusz Zelent}
\affil[1]{Institute of Spintronics and Quantum Information, Faculty of Physics, Adam Mickiewicz University, Poznan, Poland}

\author[1]{Mathieu Moalic}
\author[2,3]{Olav Hellwig}
\affil[2]{Institute of Ion Beam Physics and Materials Research, Helmholtz-Zentrum
Dresden-Rossendorf, Bautzner Landstrasse 400, 01328 Dresden, Germany}
\affil[3]{Institute of Physics, Chemnitz University of Technology, Reichenhainer Straße 70, 09126 Chemnitz, Germany}

\author[4]{Anjan Barman}
\affil[4]{Department of Condensed Matter and Materials Physics, S N Bose National Centre for Basic Sciences, Salt Lake, Block JD, Sector III, Kolkata 700106, India
}
\author[1]{Maciej Krawczyk}

\date{\today}
\begin{document}
\pagenumbering{roman}
\maketitle

\frontmatter

\tableofcontents* \newpage

\mainmatter

\begin{abstract}
    
The magnetization dynamics in nanostructures has been extensively studied in the last decades, and nanomagnetism has evolved significantly over that time, discovering new effects, developing numerous applications, and identifying promising new directions. This includes magnonics, an emerging research  field oriented on the study of spin-wave dynamics and their applications. In this context, thin ferromagnetic films with perpendicular magnetic anisotropy (PMA) offer interesting opportunities to study spin waves, in particular, due to out-of-plane magnetization in remanence or at relatively weak external magnetic fields. This is the only magnetization configuration offering isotropic in-plane spin-wave propagation within the sample plane, the forward volume magnetostatic spin-wave geometry. The isotropic dispersion relation is highly important in designing signal-processing devices, offering superior prospects for direct replicating various concepts from photonics into magnonics. Analogous to photonic or phononic crystals, which are the building blocks of optoelectronics and phononics, magnonic crystals are considered as key components in magnonics applications. Arrays of nanodots and structured ferromagnetic thin films with a periodic array of holes, popularly known as antidot lattices based on PMA multilayers have been recently studied. Novel magnonic properties related to propagating spin-wave modes, exploitation of the band gaps, and  confined modes, were demonstrated. Also, the existence of nontrivial magnonic band topologies has been shown. Moreover, the combination of PMA and Dzyaloshinskii-Moriya interaction leads to the formation of chiral magnetization states, including N\'eel domain walls, skyrmions, and skyrmionium states. This promotes the multilayers with PMA as an interesting topic for magnonics and this chapter reviews the background and attempts to provide future perspectives in this research field. 
\end{abstract}
\newpage
\setcounter{secnumdepth}{1}
\chapter{Introduction}

Magnetization textures, such as domain walls, non-uniform domain arrangements, and skyrmions are ubiquitous in magnetic materials~\cite{Chen1997NonuniformRibbons, Tatara2008MicroscopicDynamics,Malozemoff1979MagneticMaterials,Fert2013SkyrmionsTrack,mruczkiewicz2021roadmap,Fert2017MagneticApplications}. The dynamics of magnetization, where we study the time-dependent variation of magnetization, rather than its behaviour at equilibrium, is fundamental to the properties of magnetic materials, and also for respective applications. Bloch predicted a disturbance in the local magnetic order, which can propagate in a magnetic material in the form of waves in 1930~\cite{Bloch1930,Keffer1966}. It is called a spin wave (SW) since it is related to a collective excitation of spins in a ferromagnetic media, and the quantum of SWs is referred to as magnon~\cite{KittelIntroductionPhysics}. Recent interest in the use of magnetization textures as active elements in magnonic devices has been sparked by their unique properties, such as their reconfigurability, stability, and scalability toward nanoscale dimensions~\cite{Mondal2020Spin-textureDots,Streubel2015MagnetizationTextures,Albisetti2018NanoscaleSpin-textures,Wagner2016MagneticNanochannels,Chumak2022AdvancesComputing}. By utilizing magnetization textures within the framework of magnonics, research focuses on carrying and processing information using SWs, which makes it possible to harness a very rich phenomenology to designing new functionalities~\cite{Nikitov2015,Pirro2021AdvancesMagnonics}. Magnonics has many advantages that are useful in data processing, such as scalability to atomic dimensions, compatibility with complementary metal-oxide-semiconductor (CMOS) and spintronic technologies, active frequency range from a few GHz to hundreds of THz, low energy consumption, and the ability to process data over a wide range of temperatures from ultra-low to room temperature~\cite{Pirro2021AdvancesMagnonics,Chumak2022AdvancesComputing,Barman2021TheRoadmap}. 

In the late twentieth century, magnetic materials research began to focus on thin fillms magnetized out-of-plane, i.e., with perpendicular magnetic anisotropy (PMA)~\cite{Lee1997PerpendicularSystem,Nakajima1998PerpendicularStudy,Tong1998TheMultilayers,Lanchava1998MagneticFilms}.
The PMA can be artificially induced by tailoring spin–orbit coupling at the materials' interface. This is usually done by layering ferromagnets with heavy metals (for example, Pd or Pt). 
Such interfaces provide a wealth of various physical phenomena and play an important role in many effects occurring at various timescales~\cite{Hellman2017}, from ultrafast demagnetization processes to SW dynamics~\cite{Kuiper2011NonlocalLimit,Pan2018ControlledAnisotropy,Pan2022MechanismMultilayers}.
Researchers have typically focused on magnets, where the PMA is sufficiently strong so as to overcome the demagnetizing energy. The strong demand for such material properties was motivated by the desire to find a solution to increase the density of information storage and resulted in its drastic improvements~\cite{Sander2017TheRoadmap}. Furthermore, the PMA of multilayer films is of great technical interest due to their compatibility with application friendly substrates and the ability to stabilize small PMA grains in a non magnetic matrix for increasing the density of data storage on hard disks. Furthermore very narrow and robust up and down magnetic domains may also be crucial for the realization of racetrack type memory applications~\cite{Parkin2008MagneticMemory,Gaidis2017MagneticMemory,Zhao2012MagneticStorage,Ummelen2017RacetrackPinning,Kang2018ASkyrmion}.

Generally, due to the shape anisotropy, ferromagnetic thin films prefer to remain in the single domain in-plane magnetized state if the film thickness is below a critical thickness ($t_{c}$). Therefore, the $t_{c}$ parameter depends on both the PMA/demagnetizing energy ratio, the so-called quality factor $Q$, and the exchange constant. The quality factor $Q=K_{\mathrm{PMA}} / (\mu_{0}M_{\mathrm{s}}^{2}/2)$ depends on the anisotropy constant $K_{\mathrm{PMA}}$ and the magnetization saturation $M_{\mathrm{s}}$. In remanence for ultra-thin films, if $Q > 1$ then there are perpendicular magnetization domains with narrow domain walls, whereas for $Q < 1$, the film will be in-plane magnetized~\cite{Bruno1993,Kulkarni2017PerpendicularFilms}. This means that PMA materials can have complex magnetic textures, including inhomogeneities in 2D and 3D, with non-trivial demagnetizing field distributions, which makes spin dynamics in such systems an intriguing and yet interesting research problem. 

The Dzyaloshinskii-Moriya interaction (DMI) makes an important contribution to the magnetic energy of ferromagnets with contributions from spin-orbit coupling and broken inversion symmetry of crystallographic or geometrical origin~\cite{Dzyaloshinsky1958AAntiferromagnetics,Tokura2010,Soumyanarayanan2016}. In particular, it leads to the emergence of noncollinear magnetic states. It favours orthogonal spin orientation, unlike symmetric exchange coupling, as represented in the Hamiltonian by the cross product between neighbouring spins:
\begin{equation}
    \mathrm{H}_{\mathrm{DMI},\textit{ij}} = -\textbf{D}_\mathrm{DMI,\mathrm{ij}} \cdot (\textbf{S}_{i} \times \textbf{S}_{j}), \label{Eq:DMI1}
\end{equation}
where $\textbf{D}_\mathrm{DMI}$ is the Dzyaloshinskii-Moriya vector. The sign and magnitude of the DMI vector is determined by the materials involved and their order of stacking. It was shown that DMI significantly impacts the stability of noncollinear states, including formation of the N\'eel domain walls, spin spirals, skyrmions, skyrmion lattices (SkLs) and other chiral textures, offering also new application possibilities~\cite{Fert2013SkyrmionsTrack,Fert2017MagneticApplications,Muhlbauer2009SkyrmionMagnetc,Zhou2015DynamicallySkyrmions,Gobel2021}.
In particular, interfacial DMI and PMA in magnetic multilayers stabilize nanosize skyrmions, which are promising for the next generation of data storage devices, like racetrack memories~\cite{Tomasello2014AMemories}. 
Because the noncolinear and chiral magnetic textures are abundant, they also create an intriguing environment for spin dynamics and SW propagation~\cite{DosSantos2020NonreciprocityInteraction,Finco2021ImagingRelaxometry,Lonsky2020DynamicTextures}. One strategy to fabricate nanostructures with PMA and DMI based on conventional ferromagnets, is to introduce a heavy metal adjacent layer with a large spin-orbit coupling (SOC). Therefore, heavy metals, such as Pt, W, Ta and Ir, are usually used to induce the interfacial DMI, and Pd, Ta, W, Au, Pt for PMA in multilayer stacks~\cite{Magni2022KeyExpansion,Rohart2013SkyrmionInteraction,Sampaio2013NucleationNanostructures,Chaurasiya2018DependenceScattering,Chaurasiya2016DirectThickness}. In such sandwich structures, the thickness of the individual, magnetic and non-magnetic layers~\cite{Tacchi2017InterfacialThickness}, the number of repetitions~\cite{Nishimura2021InterfacialMultilayers,Pollard2017ObservationMicroscopy,Moreau-Luchaire2016AdditiveTemperature,Hrabec2014MeasuringFilms}, diffusion~\cite{Won2022StrongStructure, BeLan2020EnhancedLayer,Bandiera2012EnhancementAntiferromagnets}, as well as ion bombardment~\cite{Matczak2014TailoringAnisotropy}, play key roles in controlling the strength of these two parameters. 

In addition to the magnetization noncollinearity, the dynamics of magnetization can be also influenced by the  geometry of the sample and its spatial pattern~\cite{Streubel2015MagnetizationTextures,Yu2021}. The interplay of the magnetic properties of ferromagnets with their geometry or topology is one of the basic properties used in magnonic research. In the past few years, the SW dynamics in structured ferromagnetic materials and nonuniform magnetic configurations have been extensively investigated~\cite{Krawczyk2014ReviewStructure,De2021MagnonicGeometry,Barman2020MagnetizationPerspective,mruczkiewicz2021roadmap,Gubbiotti2019Three-DimensionalMagnonics}, which include both in-plane~\cite{Tahir2017MagnetizationLattices,Zelent2017GeometricalSpectra,Gro2021PhaseLattices,Venkat2014MicromagneticDefect,Semenova2013SpinFilter} and out-of-plane configurations~\cite{Pan2020EdgeAnisotropy,Mallick2018RelaxationAnisotropy,Saha2019FormationAnisotropy,Grafe2015PerpendicularLattices} but the latter has been very rarely studied. 
For controlling the flow of SWs, magnonic crystals~\cite{Krawczyk2001,Gulyaev2001MagnonicStructures}, which are artificial magnetic media with periodic modulation of magnetic properties, consisting of periodically ordered magnetic stripes, dots or antidots (called antidot lattices, ADLs)~\cite{Gro2021PhaseLattices,Tahir2016MagnetizationLattices,Zelent2017GeometricalSpectra,Pan2020EdgeAnisotropy,Feilhauer2020ControlledLattice,Saha2019FormationAnisotropy,Busel2018,Mandal2012OpticallyLattices,Lu2017IdentifyingAu111}, etched grooves, pits,  periodic variations of the internal magnetic field or periodic domain arrangements, are frequently used~\cite{Nikitin2015ACrystal,Rychy2015MagnonicNanoscale,Adams2011Long-rangeMnSi,Chumak2009ScatteringCrystal,Wang2010NanostructuredBandgaps,Mruczkiewicz2016CollectiveCrystal,Chumak2017MagnonicProcessing,Diaz2020ChiralFields,Zelent2017GeometricalSpectra}. They are of great interest in both pure SW physics and application-oriented magnonics. This is, because a characteristic feature of magnonic crystal is the formation of SW bands separated by the bandgaps~\cite{Krawczyk2014ReviewStructure}, analogous to the photonic and electronic band structure in photonic crystals and semiconductors, respectively~\cite{Puszkarski2003,Joannopoulos2011}. In this context, periodic magnetization textures present in thin films and the combination of magnetization textures with spatial patterning are promising approaches for advancing magnonic systems, as they can be used to excite, transmit, and process SW signals as well as perform logic operations using SWs~\cite{Pirro2021AdvancesMagnonics,mruczkiewicz2021roadmap,Sander2017TheRoadmap,Chumak2022AdvancesComputing,Mahmoud2020IntroductionComputing}.  

The combination of all these material properties, structuring and magnetization textures leads to rich physical phenomena and high application potential of spin dynamics in magnetic multilayeres with PMA~\cite{Yang2021ChiralSpintronics,Fert2017MagneticApplications,Chumak2022AdvancesComputing}.
In this chapter, we review selected studies showing important achievements in the physics of magnetization dynamics in structures based on ferromagnetic films with PMA and discuss the progress in this topic. 
The next section, Sec.~2, is dedicated to reviewing SW dynamics in homogeneously magnetized thin films and elements of confined geometry with a focus on out-of-the-plane magnetized configuration. Section~3 deals with SWs in magnonic crystals, in particular in ADLs and arrays of dots magnetized perpendicular to the film plane, emphasizing the influence of anisotropy  on the SW band structure. In Sec.~4, we discuss dynamic aspects of complex magnetic textures, with a focus on the  skyrmions in ferromagnetic materials. Finally, in Sec.~5
we discuss the perspectives of spin dynamics in PMA films for information processing applications and possible directions for further research, and finnaly we summarize the chapater.

\newpage

\chapter{Spin waves in thin films with PMA
\label{Sec:thin_films}}
To discuss spin dynamics in ferromagnetic materials with PMA we start with an introduction to SW dynamics in homogeneously magnetized ferromagnetic films in two configurations, magnetized out-of-plane and in-plane with PMA, by the external magnetic field oriented parallel, and perpendicular to the anisotropy axis, respectively.


\subsection{Ferromagnetic film magnetized along out-of-plane direction}

\begin{figure}[h]
    \includegraphics[width=\textwidth]{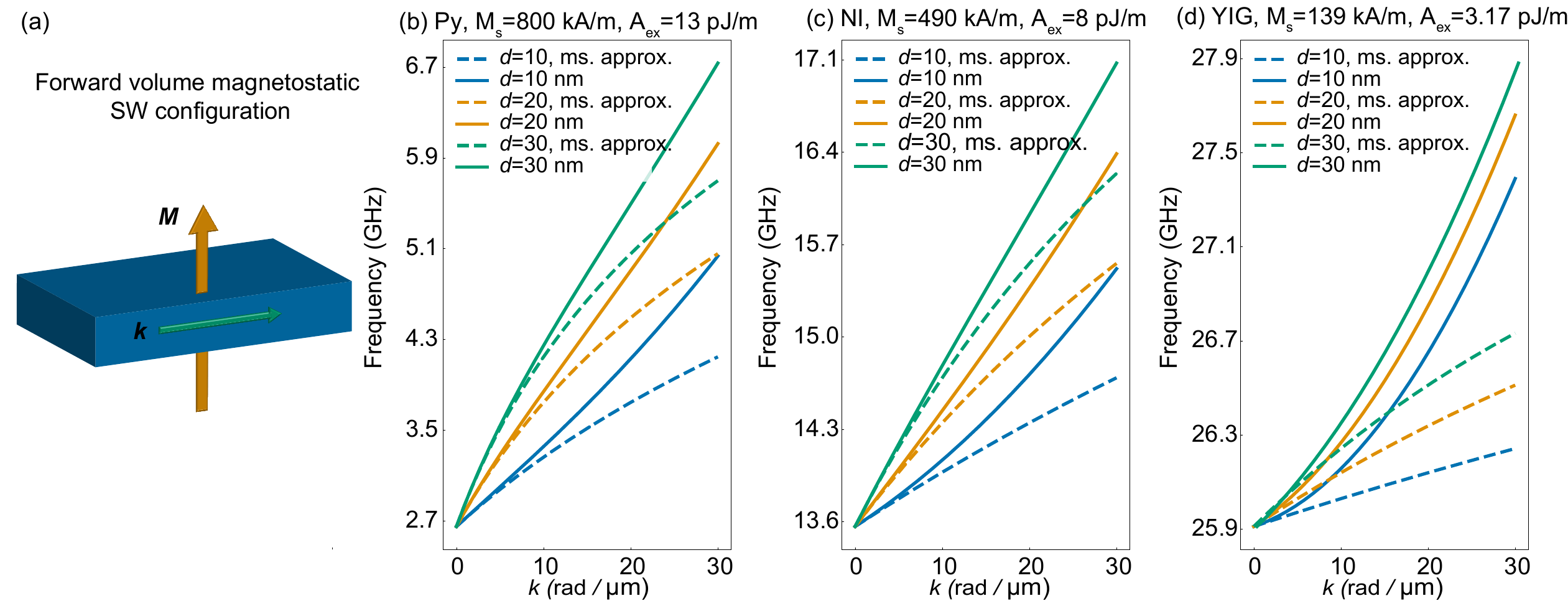}
    \centering
    \caption{ (a) Geometry of the forward volume magnetostatic SW configuration: the SWs propagate in the film plane, which  is magnetized out-of-plane. (b-d) Calculated analytically dispersion relation of SWs in a forward volume wave configuration for Py, Ni and YIG films based on Eq.~\ref{Eq:Dispersion} in external magnetic field $\mu_0 H_0=1.1$ T directed out-of-plane. Dispersions are presented for three film thicknesses 10, 20 and 30 nm and based on full Eq.~\protect\ref{Eq:Dispersion} and in magnetostatic approximation. Magnetic parameters for Ni were taken from Ref.~\protect\cite{coey2010magnetism}, Py and YIG from Ref.~\protect\cite{book:112888}. The anisotropy field was neglected in all cases.}
    \label{fig:MSFVW} 
\end{figure}

When the magnetization is parallel to the anisotropy axis, the out-of-plane direction in the case of PMA, the anisotropy field $\textbf{H}_{\text{ani}}$ has the same effect on the SW waves as the external magnetic field $\textbf{H}_0$ collinear with that axis.  The configuration, when SWs propagate in the film plane and the film is magnetized out-of-plane, is called a forward volume magnetostatic SW geometry~\cite{book:112888} (see, Fig.~\ref{fig:MSFVW} (a)). First, the analysis of SW was  considered in the magnetostatic approximation, i.e., for long wavelength SWs, with the exchange interactions neglected~\cite{Damon1965,book:112888}. Later on, the perturbative approach with exchange interactions included has been developed, and the approximate, very useful and often used, formula for the dispersion relation for the lowest-order SW mode, i.e., the mode which has homogeneous amplitude across the film thickness, has been derived~\cite{Kalinikos1980,Kalinikos1986}:
\begin{equation}
   \omega^{2} = \omega_{0} \left[\omega_{0} + |\gamma|\mu_0 M_{\mathrm{s}}\left( 1- \frac{1-\mathrm{e}^{-kd}}{kd}\right)\right ], 
   \label{Eq:Dispersion}
\end{equation}
where 
\begin{equation}
\omega_{0}=|\gamma|\mu_{0} (H_{0} + H_{\mathrm{dem}}+ {H}_{\mathrm{ani}} + H_{\text{ex}}),    
\end{equation}
 and $\omega$ is the angular frequency of SWs, $\gamma$ is the gyromagnetic ratio, $\mu_0$ the permeability of vacuum, $M_{\mathrm{s}}$ is the saturation magnetization, $k$ is the in-plane SW wavenumber, and $d$ is the film thickness. Demagnetizing field, $H_{\mathrm{dem}}$, in the case of infinite saturated ferromagnetic film, is homogeneous and equal to the saturation magnetization: $H_{\mathrm{dem}}=- M_{\mathrm{s}}$.
 $H_{\text{ex}}=\frac{2A}{\mu_0 M_{\mathrm{s}}} k^2$ is the term describing exchange interactions, with $A$ the exchange constant.
 The dispersion relation in the magnetostatic approximation is obtained by putting $H_\text{ex}=0$.  The dispersion relation of SWs in thin ferromagnetic films of different thicknesses, for both full Eq.~(\ref{Eq:Dispersion}) and in magnetostatic approximation, and three common ferromagnetic materials, Ni$_{80}$Fe$_{20}$ (Py), Ni, yttrium iron garnet (YIG), are shown in Fig.~\ref{fig:MSFVW} (b-d).

 The dispersion relation, Eq.~(\ref{Eq:Dispersion}), depends only on the in-plane wavevector magnitude $k$, not its orientation, thus the dispersion relation is isotropic. In the limit $kd \ll 1$, the group velocity  $v_\mathrm{g}(kd \rightarrow 0)\approx \frac{1}{4}\omega_\mathrm{M}d$, which is independent of frequency. It also points out that the $v_\mathrm{g}$ increases when the saturation magnetization  [compare dispersion relation for Py (largest $M_{\mathrm{s}}$), Ni and YIG (smallest $M_{\mathrm{s}}$) in Fig.~\ref{fig:MSFVW} (b-d)] or the film thickness [compare the dispersion of SWs for given material and different thicknesses in Fig.~\ref{fig:MSFVW} (b-d)] increases~\cite{Yu2012HighFilm}   that is important for enhancing transmission rates in magnonic applications. 

The SW transmission measurements in a thin  Py film magnetized out-of-plane were used to demonstrate current-induced Doppler shift of SWs~\cite{Vlaminck2008}. 
This study started the  development of all-electrical SW spectroscopy based on the broad-band vector-network analyzer ferromagnetic resonance (VNA-FMR) measurements, now widely used in magnonics, also, to realize some functional magnonic elements. Later, thin YIG film with three CPWs was used to demonstrate XNOR logic gate~\cite{Goto2019ThreeFilm}. Although the magnetization of YIG is much lower than that of Py, the ultralow damping in the former material made it ideal for magnonic applications. However, relatively large external magnetic fields needed to saturate the samples in the out-of-plane direction, i.e., approximately 1 and 0.3 T, for Py and YIG-based systems, are not very desirable for applications, in particular for integrated circuits. This obstacle can be avoided if the materials with PMA are used. Here, YIG, which is an insulator having by far the lowest SW-damping, possesses a favorable property that doping can create sufficiently strong PMA to the extent of overcoming the demagnetizing field. The PMA in garnets doped with Mn, Ga, and Bi have already been demonstrated~\cite{Chen2019SpinAnisotropy,Carmiggelt2021,Li2019}, while preserving low damping value~\cite{Soumach2018}. 
Moreover, the lattice constant mismatch between very thin YIG and a properly selected substrate can give rise to an interfacial strain as well as PMA~\cite{Ding2020}. Other rare-earth iron garnets, those based on Dy, Tm, Tb, or Eu, were also investigated and the out-of-plane  anisotropy originating from epitaxial lattice mismatch was identified~\cite{Bauer2020,Crossley2019,Rosenberg2018,Ortiz2018}.   So far, the velocity of 2.25 km/s, with the relaxation time  of 50 ns and the decay length of 115 $\mu$m (at 0.03 T external field) have been attained in 1 $\mu$m thick Bi substituted YIG film~\cite{Rao2021}. Unfortunately, the PMA introduced by doping  further reduced its pre-existing low saturation magnetization~\cite{Hansen1974}, and hence the group velocity. Thus, new materials and new ideas need to be developed to obtain thin layers with PMA, high saturation magnetization and low attenuation.
Recently, a very interesting concept as an alternative solution for the above-defined problem has been presented. It proposes to use a heterostructure composed of an underlayer with high PMA having a thin soft ferromagnetic film deposited on the top~\cite{Haldar2017}. Here, the exchange and dipolar coupling between subsystems force the out-of-plane orientation in the soft magnet, which is a media  suitable for SW propagation.

\begin{figure}[h]
    \includegraphics[width=\textwidth]{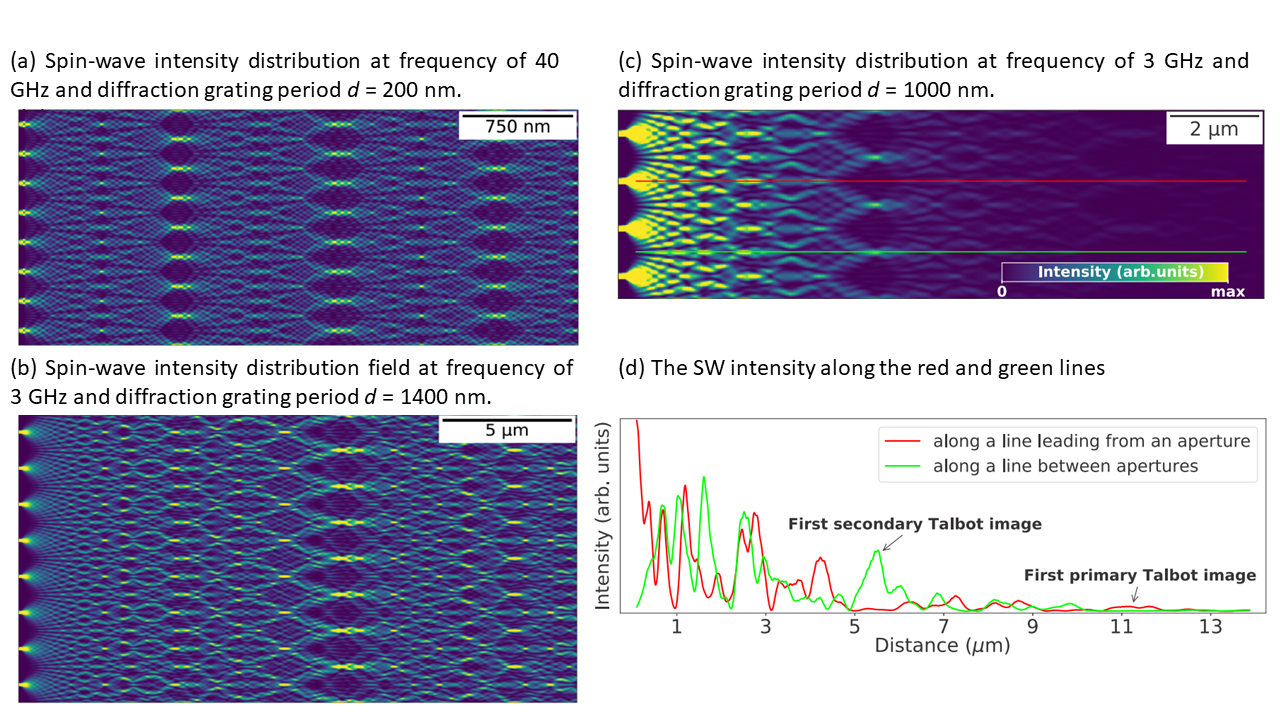}
    \centering
    \caption{ (a-b) Simulated Talbot carpet for SWs obtained for two different SW frequencies, aperture widths and diffraction grating constants in thin Py films with damping neglected. The grating is located at the left edge of each plot. (c) The SW Talbot carpet in the Py film with the damping included. The SW intensity along the red and green lines are shown in (d). Figure reproduced with permission from~\fullcite{Goebiewski2020Spin-waveFilm}. Copyright (2020) by the American Physical Society.
    }
    \label{fig:Talbot} 
\end{figure}

Another interesting characteristic of the out-of-plane magnetized films, as already mentioned is an isotropic dispersion relation, which makes the SW forward geometry identical to the electromagnetic waves in dielectrics, and makes it possible to implement magnonic analogues of many wave phenomena known from optics and photonics.
One example is the Talbot effect. It is a self-imaging process of formation of the wave-interference images of the grating at regular distances away from the grating. It was, for the first time observed for light, by H. F. Talbot in 1836 and thereon, such an interference pattern is called a Talbot carpet. The Talbot effect for SW was numerically demonstrated recently for the forward volume geometry, showing the clear formation of the Talbot carpet by SWs in thin ferromagnetic film, as shown in Fig.~\ref{fig:Talbot}~\cite{Goebiewski2020Spin-waveFilm}. As the period of the grating increases, the Talbot length, i.e., the distance between the bright focal points, also increases (Fig.~\ref{fig:Talbot}(a-b)). It is predicted that even with the reasonably large damping of Py, the secondary Talbot image can be observed (Fig.~\ref{fig:Talbot}(c-d)). However, the realization of the SW Talbot carpets in the in-plane magnetized film is disturbed by the strongly anisotropic dispersion relation of SWs in this configuration. 

The wave beams are widely explored in photonics, while in magnonics the SW optics concept has just been introduced. Here, the forward geometry ensuring isotropic dispersion relation, is also exploited. The inductive method for SW beam generation was proposed by Gruszecki \textit{et al.} in 2016~\cite{Gruszecki2016}. First, it was confirmed experimentally for the in-plane magnetized film~\cite{Korner2017}, followed by the out-of-plane configuration~\cite{Loayza2018FresnelWaves}, where additionally the Fresnel type of diffraction was observed for SWs. Just recently, also the SW birefringence has been experimentally demonstrated~\cite{Hioki2022}.

These findings demonstrate a strong analogy between the  SWs  and the basic concepts of optics and therefore pave the way for future studies of SW beam interference, which could find applications for magnonic logic devices. Indeed, the configuration with isotropic dispersion was used to demonstrate the magnonic spectrum analyzer based on the interference pattern formation~\cite{Papp2017NanoscaleInterference} and programmable SW lookup tables~\cite{Goebiewski2022Self-ImagingTables}.

\subsection{Ferromagnetic thin films magnetized in the film plane}

Ferromagnetic thin films with PMA can be in-plane saturated by applying an external magnetic field, which together with the demagnetizing field exceeds the out-of-plane anisotropy. The configuration, when the SW propagation is perpendicular to the magnetization direction (Damon-Eshbach (DE) configuration), is widely used for the determination of the DMI strength in Brillouin light scattering measurements~\cite{Di2015,Tacchi2017InterfacialThickness}. Here, the PMA in the dispersion relation is usually combined with the magnetization saturation as an effective magnetization. This approach is strictly valid under ferromagnetic resonance (FMR) conditions but not for $k \neq 0$, especially at large anisotropy values. Recent studies show that the PMA in DE configuration may result in an additional dispersion minimum at nonzero $k$ that with mode softening becomes a global minimum~\cite{Banerjee2017MagnonicSimulations}. This correlates with the phase transition to the stripe domain formation and appearing the Goldstone and Higgs SW modes~\cite{grassi2022higgs}.

\subsection{Spin-wave modes in confined geometry with out-of-plane magnetization}

\begin{figure}[h]
    \includegraphics[width=\textwidth]{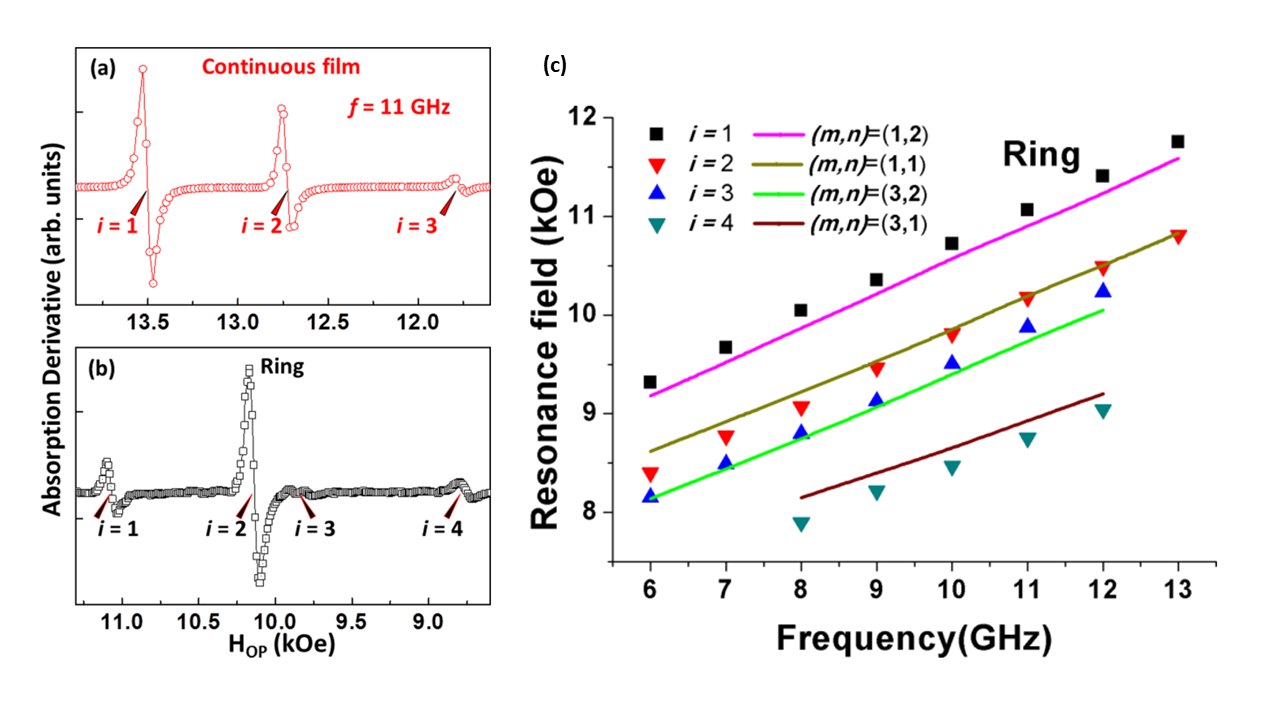}
    \centering
    \caption{ (a), (b) The ferromagnetic resonance spectra at 11 GHz from the 100 nm thick Py film and Py circular ring (of the inner 1.1 and outer 1.5 $\mu$m radius), respectively, magnetized out-of-plane. (c) Resonance fields extracted as a function of the excitation frequency for the circular ring: triangles and squares are experimental results; solid lines are theoretical calculations based on the dispersion relation, Eq.~(\ref{Eq:Dispersion}) with an inhomogeneous $H_{\textrm{dem}}$. The indices $m$, and $n$ indicate the quantization numbers of the modes along the azimuthal and radial direction, respectively. Figure reproduced with permission from ~\fullcite{Zhou2021}. Copyright (2021) by the American Physical Society.
    }
    \label{fig:FMR} 
\end{figure}

The SW spectra in confined geometry, a planar dot, with the out-of-plane magnetization is a derivative of the dispersion relation of the forward volume waves (Eq.~\ref{Eq:Dispersion}) with the discrete values of the in-plane wavenumber. Thus, the shape of the modes should preserve the symmetry of the dot shape, similarly to photonics, and hence it imposes the quantization rules on the in-plane wavevectors. However, for SWs in ferromagnetic systems, dipolar interactions  additionally influence the selection rules of the wavevector~\cite{Damon1965,Kakazei2004}. In this case, the demagnetizing field $H_{\mathrm{dem}}$ in Eq.~(\ref{Eq:Dispersion}) is inhomogeneous and mode dependent. For a dot of regular shape, the approximate formula for the demagnetizing field can be derived analytically~\cite{Kakazei2004}, and in recent literature, we can find experimental validation of this approach, see, e.g., Fig.~\ref{fig:FMR}. The SW resonances obtained in the FMR measurements, i.e. under excitation with the homogeneous microwave field, in the out-of-plane magnetized Py dots of the circular~\cite{Kakazei2004}, triangular shape~\cite{Kharlan2019StandingDots}, and rings~\cite{Zhou2021} where explained with the use of dispersion relation of SWs in thin film. 

Due to limiting sensitivity of the FMR measurements, most of the studies were performed with the array of dots separated by a distance at which the magnetostatic coupling between the dots can be neglected. Only recently, the SW resonances in the relatively thick CoFe nano-disc magnetized out-of-plane were determined from the single dot measurements~\cite{Dobrovolskiy2020Spin-waveNanodisks}. This proves the possibility of obtaining very high sensitivity in FMR measurements, especially that as many as 9 SW resonances were detected, which promises large usefulness of this measurement technique for magnonics in nanoscale elements.
\chapter{Spin-wave dynamics in magnonic crystals} \label{Sec:ADL}

As mentioned in the introduction, in recent years, magnonic crystals have received significant attention due to their potential applications in both fundamental research on linear and nonlinear wave dynamics, and in signal processing in the microwave frequency range~\cite{Krawczyk2014ReviewStructure}. These structures, having both in-plane and out-of-plane magnetization configurations, exhibit useful features, such as band folding and band gaps, in which SWs are unable to propagate, making their spectra significantly different from those of uniform media. Magnonic crystals can take a variety of forms, such as periodic magnetic patterns (bi-component, array of dots or ADLs), periodically ordered magnetic domains, or time-dependent dynamic textures -- magnonic time crystals~\cite{Trager2021Real-SpaceCrystals,Barman2020MagnetizationPerspective}. Furthermore, the magnetization dynamics can be efficiently modulated in bi-component magnonic crystals, where the periodicity may be attributed to the double sub-lattices of two different materials~\cite{Choudhury2019AnisotropicCrystal, Gubbiotti2016CollectiveNanowires}, or the different orientation of magnetization~\cite{Pan2020EdgeAnisotropy}. 

So far, it has been shown that such structuring can be used to design magnonic sensors~\cite{Talbot2015ElectromagneticNDT,Takagi2014MonolithicCrystals}, filters~\cite{Sadovnikov2018Spin-WaveCrystals,Merbouche2021FrequencyFilms}, amplifiers~\cite{Kumar2018ResonantCavity}, phase shifters~\cite{Hayami2021PhaseCrystals,Zhu2014MagnonicShifters,Zhang2021PhaseWalls}, couplers~\cite{Yu2013OmnidirectionalCoupler,Sadovnikov2016NonlinearCrystals}, multiplexers~\cite{Frey2020Reflection-lessCrystal}, transistors~\cite{Chumak2014MagnonProcessing} and logic gates~\cite{Chumak2014MagnonProcessing,Khitun2010MagnonicCircuits,Nikitin2015ACrystal}. Most of the proposed magnonic structures consist of one- and two-dimensional structuring based on thin ferromagnetic films. 
The combination of the advantages of the out-of-plane magnetization induced by the presence of PMA and the aforementioned properties of magnonic crystals raises great interest for further study. In particular, the ADL type and  the periodic magnetic textures have drawn intense research interest and it will be reviewed below.

\subsection{SW spectra in antidot latices}
The dynamics of SWs in PMA materials over the past few decades has been measured mainly in thin films~\cite{Pal2011TunableThickness,Pal2012Time-resolvedSystems}. Then with the development of patterning methods and the manufacture of multilayer materials, theoretical and experimental research was carried out on various materials and different patterning~\cite{Bali2012High-symmetryPlane,Schwarze2015UniversalMagnets,Grafe2016GeometricAnisotropy,Mandal2012OpticallyLattices,Pal2014,Pan2020EdgeAnisotropy}. 

\begin{figure}
    \includegraphics[width=1.0\textwidth]{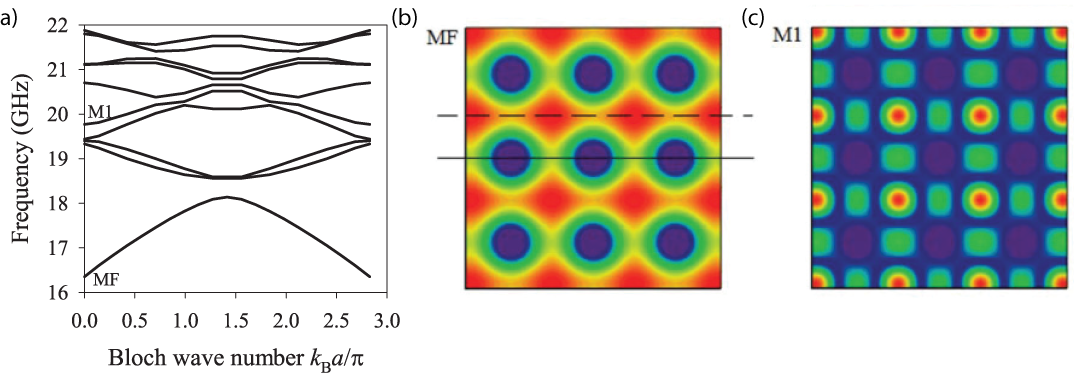}
    \caption{(a) Dispersion relation of SWs along the [11] crystallographic direction in the ADL based on the square array of holes (lattice constant $a=415$ nm) in the 40-nm-thick Py film. The static magnetic field is 1.5 T and is applied perpendicular to the film surface. (b-c) Calculated SW-mode profiles for the modes MF and M1 marked in (a). Figure reproduced with permission from \fullcite{Bali2012High-symmetryPlane}. Copyright (2012) by the American Physical Society.
    }
    \label{fig:BALI_ADL}
\end{figure}

For example, FMR spectrum for perpendicularly magnetized ADLs based on layers made of Py and square lattice of circular holes has been studied both experimentally and theoretically~\cite{Bali2012High-symmetryPlane,Schwarze2015UniversalMagnets,Grafe2016GeometricAnisotropy}. Bali \textit{et al.} in Ref.~\cite{Bali2012High-symmetryPlane} measured resonance frequencies using microstrip FMR in ADL of 415 nm period. A number of modes (see, Fig.~\ref{fig:BALI_ADL} (a)) have been found, including the fundamental mode of ADL with the lowest frequency (yet greater than the FMR frequency of a uniform Py film), and higher-order modes with frequencies increasing with the quantization number (see Fig. Figure~\ref{fig:BALI_ADL} (b-c)). Frequencies of all these modes linearly depend on the perpendicularly applied magnetic field value. Although those measurements didn't allow to determine of the dispersion relation, they computed it numerically for the two directions of SW propagation, along the main axis of the square lattice and along the diagonal. They found the bands splitting but their frequency positions and widths depended on the propagation direction, which makes it difficult to open a full magnonic bandgap. This study also revealed that for the diagonal direction of propagation, as the hole diameter increases, the width of the first band decreases, whereas the band separation increases.

The subject of SWs in perpendicularly magnetized square-lattice ADLs made of Py was also considered by Schwarze \textit{et al.} in Ref.~\cite{Schwarze2015UniversalMagnets}. It was shown using micromagnetic simulations that in the ADL based on a 22 nm thick film and circular holes of a diameter 120 nm, a full bandgap is present. An increase of both the SW band frequencies and the width of the bandgap with decreasing area of the ferromagnetic material were shown in both papers cited above. These show a possibility for optimization of the magnonic band structure by an appropriate design of the magnonic crystal geometry.  

To saturate an ADL along the out-of-plane direction sufficiently strong magnetic field has to be used, which is not suitable for miniaturization and integration. Therefore, ADLs made of thin-film systems with PMA are particularly promising for magnonic crystal applications. 
Example systems with strong PMA are Co/Pd multilayers. Such systems, with focused ion beam (FIB) milled holes of diameter 100 nm obtained by Ga-ion irradiation, were studied by Pal \textit{et al.} in Ref.~\cite{Pal2014}. 
All-optical measurements of SW dynamics with time-resolved magneto-optical Kerr effect microscopy  revealed how the SW resonance frequencies depend on the lattice constant.
Analogous to the case of out-of-plane magnetized ADLs in Py, the frequencies decrease with increasing separation between the holes. However, here, for some resonant modes, an opposite tendency was also observed. 
Interestingly, it was shown that these modes may have frequencies lower than the FMR frequency of the uniform Co/Pd multilayer (without holes). Theoretical analysis using the plane wave method (PWM) revealed that the low-frequency modes can be associated with magnetization oscillations in the vicinity of holes, and they were called shell modes (edge modes). It was proposed that these modes emerge due to locally, near the edges of the holes, induced modifications in the material parameters of the Co/Pd multilayer  by the Ga bombardment. In the calculations, a reduction of the values of magnetocrystalline anisotropy, effective magnetization, and exchange constant was assumed. Although the local reduction of the PMA could also cause the in-plane rotation of magnetization in the holes' shells at remanence, this effect was not taken into account  until  recently~\cite{Pan2020EdgeAnisotropy}. 
\begin{figure}
    \includegraphics[width=1\textwidth]{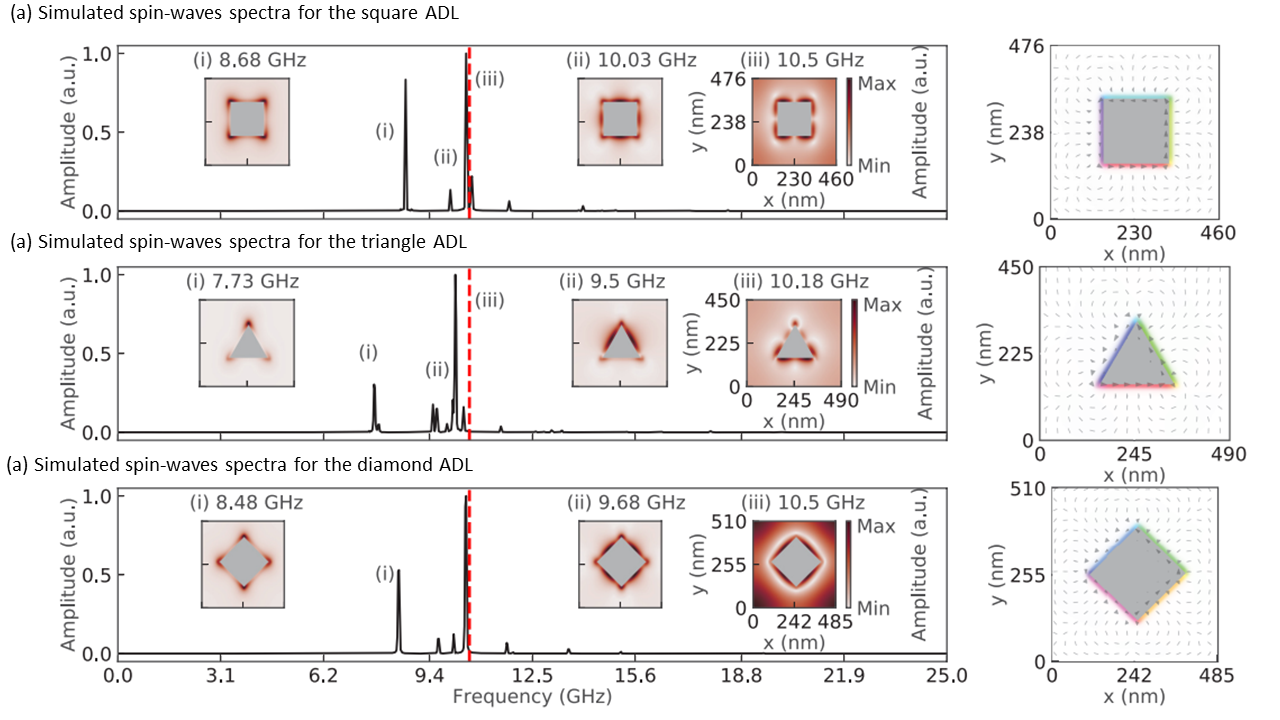}
    \caption{
    Simulated SWs spectra for the square (a), triangle (b) and diamond ADL (c) structures with linear hole size $d$ = 10 nm, the respective static magnetization configurations and the profiles of the most intense lines from the spectra. The simulations are performed with an out-of-plane magnetic field $H$ = 2.23 kOe, but the SW spectra are calculated from the in-plane component of the magnetization.
    Figure reproduced with permission from     \fullcite{Pan2020EdgeAnisotropy}. Copyright (2022) by the American Physical Society.
    }
    \label{fig:pan_ADL}
\end{figure}

The recent study by Pan \textit{et al.} in Ref.~\cite{Pan2020EdgeAnisotropy} continues the topic of SWs in the ADL lattice  based on Co/Pd multilayer  with PMA. Here, the research was focused on the influence of the magnetization orientation in the shells on the SW dynamics (Fig.~\ref{fig:pan_ADL} a-c)  in unsaturated state.  
This study was focused on the square-lattice ADL with different shapes of the holes. It was also assumed that the shells around holes have significantly reduced PMA,  due to the Ga+ ion irradiation during the patterning process. At remanence and small magnetic fields, in these areas the demagnetizing energy prevail, creating areas with complex (vortex or onion-like) magnetic textures around the holes (see the right panel in Fig. \ref{fig:pan_ADL} a-c). In these regions of width down to 5 nm, the magnetization stabilizes in the film plane forming suitable conditions for localization of low-frequency SWs (see the left panel in Fig. \ref{fig:pan_ADL} a-c). Furthermore, it was shown that the antidots with triangular, square, and diamond antidots can have different magnetization configurations at remanence, resulting in different oscillation frequencies. Such edge-localized SWs, which are strongly influenced by geometry  and magnetization configuration, are worthy candidates for re-configurable magnonic devices. 

\subsection{SW guiding in magnonic crystals}

\begin{figure}[!htb] 
\centerline{\includegraphics[width=\textwidth]{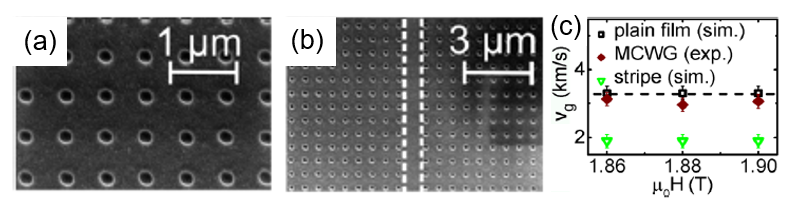}}
\caption{(a) Scanning electron microscopy image of ADL.
(b) Scanning electron microscopy image of ADL with removed one row of holes.
(c) Comparison of the group velocities of SWs in a plane film (the black squares), magnonic crystal waveguide (the brown dots) and in the stripe of the same width as SW channel in ADL (the green triangles).
Figure reproduced with permission from     \fullcite{Schwarze2013MagnonicCoFeB}. Copyright (2022) by the American Institute of Physics.}
\label{Fig:ADL} 
\end{figure}

One of the flagship uses of periodic structures in photonics is guiding electromagnetic waves. Similar studies have also been performed in magnonics for perpendicularly magnetized ADLs.
Schwarze \textit{et al.} in Ref.~\cite{Schwarze2013MagnonicCoFeB} applied this concept  to ADL made from CoFeB layer with PMA (see Fig.~\ref{Fig:ADL}(a-b)) by removing one row of holes, the part which was used to create a channel and guide SWs. Indeed, the SWs at frequencies from the bandgap of the ADL spectrum were confined to the created channel, and they allow for signal transmission. The all-electrical SW spectroscopy utilizing two coplanar waveguides was used to investigate the impact of the lattice constant on the group velocity of SWs. Interestingly, it revealed that it is possible to achieve higher group velocities in such magnonic crystal waveguide as opposed to an ordinary waveguide made of a single stripe of the same width as the SW channel in the ADL (see Fig.~\ref{Fig:ADL}(c)).
The topic of magnonic crystal waveguides based on the out-of-plane magnetized ADLs was further conducted by Chi \textit{et al.}~\cite{Chi2014ConfinementDefects}, but it was based on YIG film saturated by the external magnetic field. 

\subsection{Topologically protected propagating edge SWs} 

The other interesting property of the out-of-plane magnetized ferromagnetic elements forming periodic patterns is that they can form favorable conditions  for existence of the edge magnonic bands with topological protection. 
This is due to the unique, as compared to photonics and phononics, property of magnonic systems, which is the presence of chiral interactions: the ubiquitous magnetostatic~\cite{Kruglyak2021} and DMI.

The existence of edge SW states were shown for bicomponent magnonic crystals  with a complex unit cell composed of YIG film hosting Fe dots~\cite{Shindou2013TopologicalCrystal}. Due to magnetostatic interactions, the  propagation of SWs from topologically protected bandgaps, along the edges of the system, is robust against irregularities in the periodicity of the lattice, boundary roughness, and it is free of any elastic backward scatterings with moderate strength.   Also, due to chiral magnetostatic coupling in a lattice of ferromagnetic discs and pillars saturated out-of-plane, the formation of topological SW edge modes were predicted~\cite{Shindou2013TopologicalCrystal,Lisenkov2014}. In Ref.~\cite{Lisenkov2014}, besides of the time-reversal symmetry breaking of the magnetostatic interactions, the inhomogeneity of the internal magnetic field at the lattice edges was also indicated as a possible mechanism for the SW edge states formations. Interestingly, not only magnetostatic interactions, but also DMI which have a chiral nature, may provide suitable conditions for formation of the topologically protected SW bands at the edges of the artificial crystals, including SkLs, which will be described later in this review~\cite{Roldan-Molina2016TopologicalCrystal,Zhang2013,Diaz2020ChiralFields}.   

\begin{figure}[!tb] 
\centerline{\includegraphics[width=15cm]{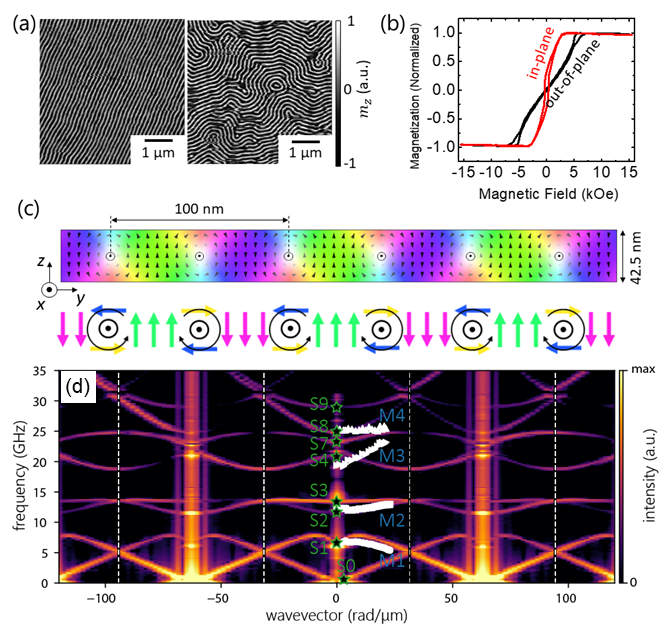}}
\caption{(Color online)
(a) MFM image revealing the parallel (left panel) and
labyrinth (right panel) weak stripe domains in Co/Pd multilayers. 
(b) Magnetic hysteresis loops of  Co/Pd multilayers with in-plane and out-of-plane applied magnetic fields. 
(c) The crossectional view through the thickness of the magnetization configuration obtained by means of micromagnetic simulations. Arrows indicate the
magnetization vector. A schematic representation of this magnetization distribution with clockwise and anticlockwise domain walls is shown in the bottom panel.
(d) Dispersion relation for the propagation of SWs across the domain walls (along the $y$-axis). The colormap in the background represents the results of micromagnetic simulations. In contrast, the white points indicate experimentally measured dispersion by BLS.
Figure reproduced with permission from 
\fullcite{Gruszecki2019ThePatterns}. 
Copyright (2022) by Elsevier Inc.}
\label{Fig:fig_stripeDomains} 
\end{figure}

\subsection{SW bands in stripe domain structure}

Magnetic texture-based magnonics is becoming an excitingly growing subfield of magnonics~\cite{Yu2021, mruczkiewicz2021roadmap, petti2022review}. The main advantages of the utilization of magnetization textures as a medium for SW propagation are the magnetic field-driven reconfigurability, so a complex nanostructuring is not required and the impact of defects may be reduced, these systems  do not require any static external magnetic field during their operation, and offer the ability to couple SWs with the magnetization texture dynamics.
One of the basic classes of systems studied so far are stripe domain patterns. Such systems can occur in media with PMA in ultra-thin films  with $Q>1$. For mono and multi- layers with $0<Q<1$ stripe domain patterns may also emerge, however, above a certain critical thickness~\cite{vukadinovic2000PRL, marko2019tunable}. 
The critical thickness above which weak stripe domain patterns with alternating out-of-plane component of magnetization appear depends on the value of $Q$. The critical thickness can range from a few tens of nanometers to values even greater than a few hundred nanometers, for materials with  $Q\ll1$.
Such textures in remanence are characterized by an effective non-zero in-plane component of magnetization resulting from the internal structure of the domain walls~\cite{hubert2008book, fin2015plane}. These domain walls are complex, non-uniform across the thickness and resembles vortices with cores aligned in one direction, see Fig.~\ref{Fig:fig_stripeDomains} (c). 

Resonance spectrum in stripe domain patterns is a subject of years-long investigation~\cite{artman1978ferromagnetic, vukadinovic2000PRL, ebels2001fmr, alvarez2004micromagnetism}. It was shown, among others, that the response of the system strongly depends on the polarization of the microwave field~\cite{vukadinovic2000PRL}. This is due to the fact that a microwave field linearly polarized along the direction of equilibrium magnetization in a particular part of the texture does not excite modes  in this region. 
Interestingly, although, the resonance spectrum in stripe domain patterns is well-known, the utilization of stripe domain patterns as magnonic crystal is a relatively new topic of research.
The first theoretical papers investigating SW propagation in stripe domains obtained in ultra-thin films with PMA in the context of their use as magnonic crystals were published  in 2015 and 2016~\cite{wang2015spin, wang2016magnonic, borys2016spin}.
Wang \textit{et al.}~\cite{wang2015spin} demonstrated analytically the band gap opening in such magnetization texture. 
Borys \textit{et al.}~\cite{borys2016spin} studied the effect of DMI interaction, which among others, changes the type of domain walls, from Bloch at zero DMI to N\'eel for large values of DMI, on the opening of band gaps in periodic stripe domains.

The primary technique for measuring the dispersion relation of SWs is Brillouin light scattering (BLS) spectroscopy, but it can also be used to measure other properties of SW dynamics like zero wavevector modes~\cite{camara2017magnetization}, modes' imaging~\cite{Sebastian2015Micro-focusedNanoscale}, or to determine the strength of DMI~\cite{dhiman2022magnetization}, for example, in stripe domain patterns~\cite{camara2017magnetization, Banerjee2017MagnonicSimulations}. Camara \textit{et al.}~\cite{camara2017magnetization} investigated how the frequency of the modes localized at the surface of the stripes domains changes with magnetic field strength. They used BLS and FMR measurements to investigate 78 nm $\alpha$-Fe film. They observed two sets of modes with different dependencies on the magnetic field, where the first has been localized at the surface of the stripes domains, while the second  at the inner region of these stripes where the local magnetization is perpendicular to the surface. Dhiman \textit{et al.}~\cite{dhiman2022magnetization} using BLS measurements on (Ir/Co/Pt)$_{6}$ multilayers and non-zero wavevector SWs, have quantitatively evaluated the interfacial DMI strength as the difference between Stokes anti-Stokes peak frequencies. The first experimentally measured dispersion relation of SWs propagating across 
a stripe domain pattern was reported by Banerjee, \textit{et al.} in Ref.~\cite{Banerjee2017MagnonicSimulations}. In this study, a Co/Pd multilayer of total thickness around 42 nm characterized by the moderate value of PMA ($Q<1$) was considered. Fig.~\ref{Fig:fig_stripeDomains} shows a schematic of the system studied in Ref.~\cite{Banerjee2017MagnonicSimulations}, the equilibrium magnetic configuration, and a magnetic force microscope (MFM) image. Aligned stripe domains with a lattice constant of about 100 nm are visible. BLS measurements detected 4 bands for propagation across the domain walls. These results agree well with the micromagnetic simulations. Moreover, the simulations reveal the opening of bandgaps. This topic was further discussed in detail in Ref.~\cite{Gruszecki2019ThePatterns}.
Interestingly, the two lowest modes originate from the oscillations of magnetization texture and can be associated as originating in the Goldstone mode. Furthermore, it was recently reported that while the stripe domain pattern emerges (for field values just below the critical field), we can observe also the occurrence of the Higgs mode~\cite{grassi2022higgs}.
 The bands of higher frequency can be identified as ordinary SWs.
The influence of the polarity and chirality of domain walls on SW dynamics was also studied in Ref.~\cite{Gruszecki2019ThePatterns}.

Another intriguing topic is the propagation of SWs along domain walls, which are considered to be a promising type of ultra-narrow waveguides~\cite{Wagner2016MagneticNanochannels,Garcia-Sanchez2015NarrowWalls}. Banerjee \textit{et al.} in Ref.~\cite{Banerjee2017MagnonicSimulations} showed  the experimentally measured dispersion relations of a few SW bands  propagating along the domain walls. Interestingly, the simulation results indicate that this propagation is unidirectional, i.e., the direction depends on the domain-wall chirality. This unidirectionality was explained by Henry \textit{et al.} in Ref.~\cite{Henry2019UnidirectionalType}.

\chapter{Spin dynamics in topologically protected spin textures} \label{Sec:textures}

As mentioned in the introduction, DMI is another, along the PMA, important interaction that leads to the emergence of noncollinear magnetic states. As DMI favors the perpendicular alignment of neighboring spins~\cite{Hrabec2014MeasuringFilms,Vida2016Domain-wallInteraction}, the simultaneous presence of the symmetric exchange leads to the formation of chiral domain walls~\cite{Legrand2018HybridMultilayers,Safeer2022EffectSkyrmions,Emori2013Current-drivenWalls}, and also skyrmions~\cite{Gobel2019ElectricalDevice,Beg2017DynamicsNanostructures,Zelent2022StabilizationNanostructures,Zelent2017Bi-StabilityInteraction,Saha2019FormationAnisotropy}. Magnetic skyrmions were first experimentally observed in bulk non-centrosymmetric crystals~\cite{Yu2010Real-spaceCrystal,Muhlbauer2009SkyrmionMagnet}, and later on in single ultrathin films~\cite{Nagaosa2013TopologicalSkyrmions}. The origin of DMI is responsible for the formation of different skyrmion configurations: in bulk DMI, so-called Bloch-type skyrmions are mainly formed, and in multilayers with heavy metal/ferromagnet interfaces, the interfacial DMI results in N\'eel-type skyrmions formation~\cite{Buttner2018TheoryApplications, Boulle2016Room-temperatureNanostructures, Beg2015GroundNanostructures,Desplat2018ThermalEigenmodes,Riveros2018AnalyticalSkyrmion}. Currently, with respect of potential applications, the research is focused on metallic heterostructures (multilayers) with thin ferromagnetic layers coupled to materials with strong spin-orbit couplings~\cite{Back2020TheRoadmap, Giustino2021TheRoadmap}. 

Skyrmion dynamic has been and continues to be studied for many different configurations, systems and material compositions. This includes isolated and multiple skyrmions in nanodot~\cite{Zelent2022StabilizationNanostructures,Mruczkiewicz2018AzimuthalSkyrmions,Rajib2021RobustDMI,Guslienko2015SkyrmionAnisotropy,Zeng2020DynamicsTorque}, as well as skyrmion chains in magnetic stripe or lattice of dots~\cite{Suess2018AMemory,Du2015Edge-mediatedGeometry,Mruczkiewicz2016CollectiveCrystal} and two-dimensional SkLs (skyrmion crystals)~\cite{Hirschberger2019SkyrmionLattice,Muhlbauer2009SkyrmionMagnet,Han2010SkyrmionMagnet,Hayami2021PhaseCrystals,Wu2021SizeCrystals,Wang2020SkyrmionGas,Yu2010Real-spaceCrystal}. The materials used in these investigations include ultra-thin ferromagnetic films and multilayered structures~\cite{Tomasello2014AMemories,Parkin2008MagneticMemory, Bessarab2018LifetimeSkyrmions}. Such a wide interest in skyrmions is justified by their potential applications. The main perspective envisaged for N\'eel skyrmions are ultra-dense memories and computing systems, in particular racetrack memories and brain-inspired computing systems~\cite{Gobel2021,Pan2019ANetwork,Li2017MagneticDevices,Vakili2021SkyrmionicsComputingMagnets}. However, the use of skyrmion still requires a few important challenges to be overcome. Selection of the compounds and manufacturing of high-quality multilayers suitable for skyrmion formation and motion, controlled and reproducible nucleation of skyrmions, control of skyrmion motion on long distances as well as excitation and detection of the skyrmion dynamics, are among them. 

There are a variety of mechanisms already identified for the nucleation of magnetic skyrmions. These include the use of magnetic~\cite{Liu2020NucleationNanostructures,Vetrova2021InvestigationDot} or electric fields~\cite{Desplat2021MechanismNucleation}, electric currents~\cite{Hrabec2017Current-inducedBilayers,Sampaio2013NucleationNanostructures,Zhao2021DeterministicCurrent}, thermal gradients~\cite{Wang2020ThermalSkyrmions}, besides skyrmion generation during a sample remagnetization process~\cite{Saha2019FormationAnisotropy,Wang2021StimulatedMagnet}, induced by magnetic force microscopy tip~\cite{Zelent2021SkyrmionTip}, 
geometrical constrictions from domain walls~\cite{Jiang2015BlowingBubbles,Pathak2021GeometricallySkyrmions} or laser heating~\cite{VinasBostrom2022MicroscopicFilms,Berruto2018Laser-inducedMicroscope,Je2018CreationFilms,Polyakov2020GenerationPulses,Koshibae2014CreationHeating}. The racetrack memories require high skyrmion velocities, so enhancing torque-induced motion of skyrmion, i.e., electric current. Here, the main obstacle is skyrmions' collision with the boundaries of the medium due to the skyrmion Hall effect~\cite{Chen2017Spin-orbitronics:Effect}.

Much less attention is devoted to spin dynamics, in particular skyrmion dynamics and skyrmion interaction with SWs. In the following sections, we provide an overview of selected studies of SW dynamics in magnetic skyrmions related to mentioned challenges.

\subsection{Nucleation and stabilization of magnetic skyrmions by SWs}

Despite substantial success in terms of skyrmion nucleation using a number of different approaches, as mentioned above, the use of the SWs and spin dynamics for skyrmion formation has not been widely explored~\cite{Liu2015SkyrmionWaves,Wang2020Spin-waveGeneration,Yao2021MagneticFocusing}. However,  such an effect can be very interesting in magnonics. 
\begin{figure}
  \includegraphics[width=1.0\textwidth]{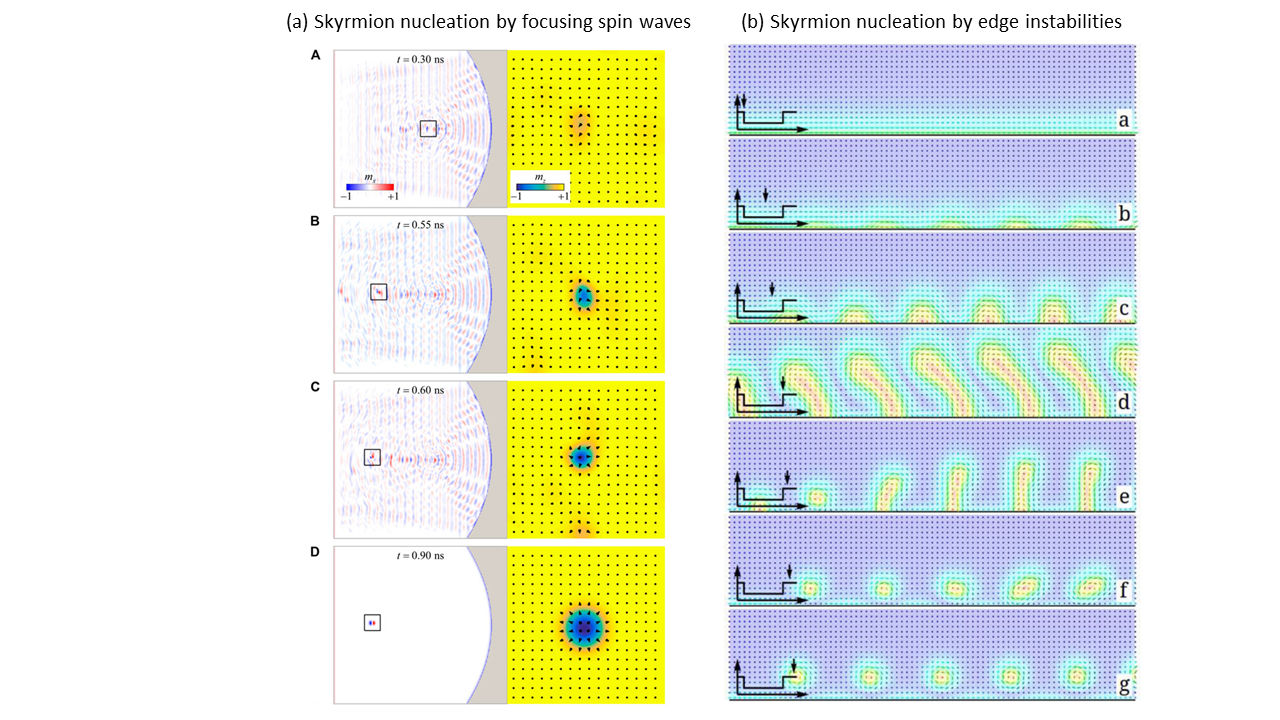}
  \caption{ (a) The process of creating the magnetic skyrmion by focusing the reflected SW beam. The exciting microwave magnetic field is applied in (A, B, C), and is turned off in (D). The $z$-component of the magnetization in the rectangular area from the left column is enlarged in the right column. Figure reproduced with permission from \fullcite{Yao2021MagneticFocusing}, under the terms of license CC BY 4.0. (b) Creation of a chain of skyrmions from the edge-localized SWs using a local instability of the magnetization at the film edge. The panels show snapshots at various times and various magnitudes of the external magnetic field, according to the schemes shown at the left-bottom corners. The colour denotes the $z$-component of the magnetization obtained by micromagnetic simulations. Figure reproduced with permission from \fullcite{Muller2016EdgeLayers} under the terms of license CC BY 4.0.}
  \label{fig:skyrmion_nucleation}
\end{figure}
Recently, Yao \textit{et al.}~\cite{Yao2021MagneticFocusing}  showed the method of generating magnetic skyrmions in the effective Co layer with DMI and PMA by focusing SWs totally reflected by a curved film edge. 
This study numerically demonstrated that the accumulation of SW energy in the focal spot can be sufficient to form or destroy a skyrmion. In this method, the SW intensity at the focal point increases significantly, showing a strong magnetization oscillation, as seen in Fig.~\ref{fig:skyrmion_nucleation} (a, subplot A). As SWs are continuously excited, more energy can be accumulated, resulting in  the switching of the magnetization locally and the formation of magnetic droplets, which can be easily driven by SWs, as shown in Fig.~\ref{fig:skyrmion_nucleation} (a, subplot B). A magnetic droplet is then transformed into a stable skyrmion under the continued influence of SWs as shown in Fig.~\ref{fig:skyrmion_nucleation} (a, subplots C-D). 

Another process, which involves SWs in the formation of skyrmions has been proposed by Muller, \textit{et al.}~\cite{Muller2016EdgeLayers}. In a thin film, a magnetization canted at the film edge, appearing due to DMI, creates favourable condition for the existence of SWs, which are bounded to the film edge, i.e., edge SWs. 
DMI introduces also nonreciprocity to the SW dynamics and minima in the dispersion relation of edge SWs for positive (or negative) wavevectors is formed. The softening of the edge mode at nonzero wavenumber with decreasing magnetic field results in magnetization instability. That way, skyrmions can be created in a controlled manner, just by using the specific spatial and time protocol of the external magnetic field. As shown in Fig.~\ref{fig:skyrmion_nucleation}(b),  the edge SWs locally destabilize the magnetization after decreasing the magnetic field, thereby initiating the formation of the merons~\cite{Ezawa2011CompactFilms, Gao2019CreationFilms} (i.e., half of the skyrmion), which grow, and move into the film area over time (see, Fig.~\ref{fig:skyrmion_nucleation}(b), subplots b-d). An increase of the magnetic field eventually leads to the formation of the skyrmions from the merons detached from the film edge (see Fig.~\ref{fig:skyrmion_nucleation}(b), subplots e-f). Although, the theory was developed for the 2D ferromagnetic system without magnetostatic interactions, it convincingly explains the experimental observation of the skyrmion chain formation in FeGe stripes~\cite{Du2015}, and the scenario should apply also to the multilayered films with the interfacial DMI.


\subsection{Eigenoscillations of the magnetic skyrmion in constrained geometries}

\begin{figure}
  \includegraphics[width=\textwidth]{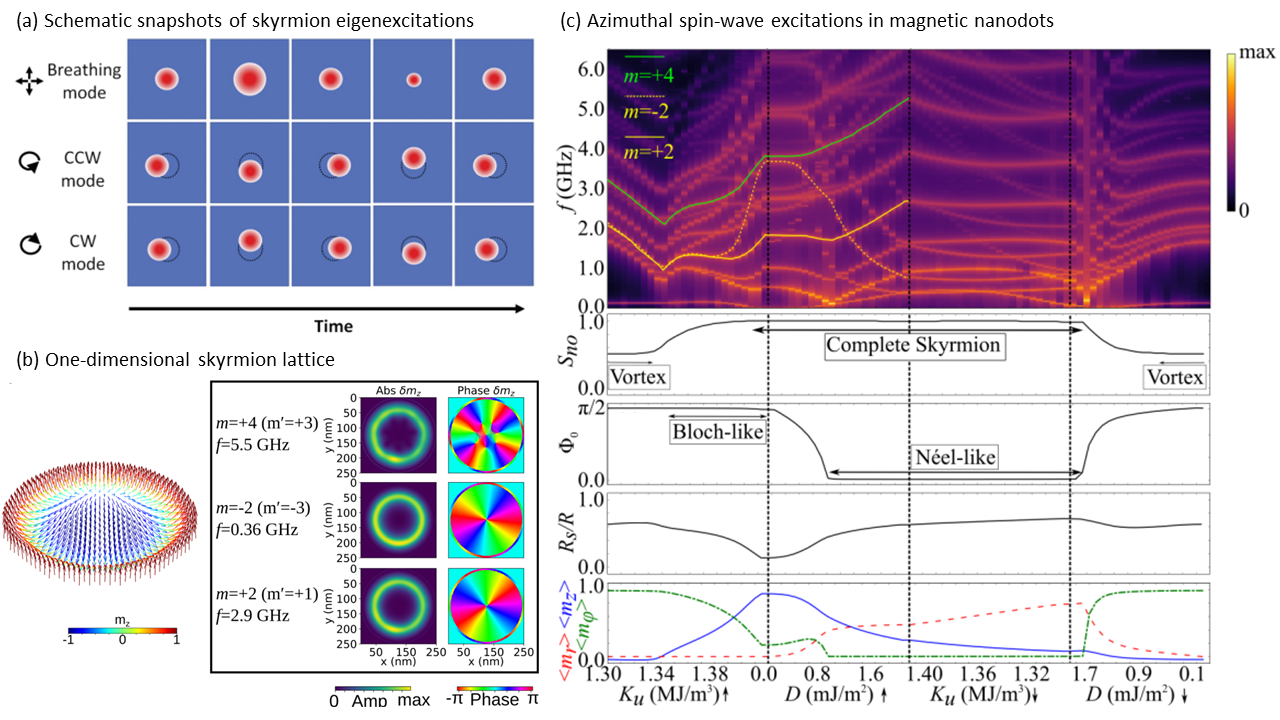}
  \caption{ (a) Schematic snapshots in different moments in time of skyrmion eigenexcitation for three types of modes: breathing mode, CCW (counter-clockwise mode) and CW clockwise gyrotropic mode. Figure reproduced from~\fullcite{Lonsky2020DynamicTextures} under the terms of license CC BY 4.0. (b) Spatial distribution of the static magnetization (left column) and SW eigenmodes (right column) (the amplitude and phase of the $z$ dynamic magnetization component) for three different frequencies with a different azimuthal number in the nanodisk with the N\'eel skyrmion. (c) Frequencies of SWs excited along the paths between four inhomogeneous magnetization configurations obtained by changing PMA constant and DMI in the ferromagnetic nanodisk: (starting from the left) vortex, Bloch and N\'eel like skyrmions, and again vortex. Below, the static properties of the respective solitons: (from the top) the skyrmion number (topological charge), the average skyrmion phase, the reduced skyrmion radius, and the average magnetization components. Figures (b) and (c) reproduced with permission from~\fullcite{Mruczkiewicz2018AzimuthalSkyrmions}. Copyright (2022) by the American Physical Society.
  }
  \label{fig:skyrmion_dynamicsa}
\end{figure}

Skyrmion dynamic, in particular excited by the microwave magnetic fields, is promising for application in the field of magnonics.
There are four characteristic types of the skyrmion modes~\cite{Mruczkiewicz2018AzimuthalSkyrmions,Gareeva2018CollectiveChain,Satywali2018GyrotropicMultilayers,Kim2014BreathingDots,Kim2018CoupledLattices}: breathing, gyrotropic, radial and azimuthal, which are partially presented in Fig.~\ref{fig:skyrmion_dynamicsa}. The modes can be classified according to the radial (\textit{n}) and azimuthal (\textit{m}) indices, which correspond to the number of radial, and azimuthal wave nodes, respectively. For the gyrotropic modes, the skyrmion center may rotate clockwise (CW) or counterclockwise (CCW) around the center position  ($n = 0$ and $m =\pm 1$) depending on the core polarization. The frequency of  the gyrotropic mode is generally between zero and 1 GHz. This mode represents a skyrmion center circular movement, and therefore it is more prominent for smaller skyrmions. The strong coupling of skyrmion gyrotropic motion with the low $m$ azimuthal modes is expected for larger skyrmions, resulting in appearance of CW and CCW gyrotropic motion for the same state. Azimuthal modes, are similar to a gyroscopic mode, but the center of the skyrmion remains in the same position, while only the SWs propagate along an azimuthal path ($n = 0$ and $m \neq 0$) (see, Fig.~\ref{fig:skyrmion_dynamicsa} (b)). For the breathing mode, the area of skyrmion extends and shrinks periodically ($n = 0$ or 1 and $m=0$). The experimental demonstrations of skyrmion dynamics are still limited due to the high damping of skyrmion hosting structures, difficulties of sample fabrication, and very selective sensitivity of the skyrmion modes on the microwave field. 

Magnetic texture strongly depends on the material parameters and external factors such as a magnetic field or electric current. Under given conditions, the magnetic skyrmion can be nucleated, annihilated or transformed into another texture like a complex domain or vortices. Such a state change can be dynamically controlled by external factors. This ability to control the magnetic texture can be used to control SW bands. Mruczkiewicz, \textit{et al.} in Ref.~\cite{Mruczkiewicz2018AzimuthalSkyrmions} studied azimuthal SW's excitations in a circular ferromagnetic nanodot in different inhomogeneous, topologically nontrivial magnetization states. In particular, they studied SW dynamics along the closed continuous path with transitions as: vortex→Bloch-type skyrmion→N\'eel-type skyrmion→vortex state. These transitions were realized by gradually changing the out-of-plane magnetic anisotropy and the DMI value. Interestingly, the CW and CCW modes of the same azimuthal number $m$ have the same frequency only in the vortex state. As soon as the magnetization rotates from the in-plane alignment of the vortex state, the CW and CCW modes attain different frequencies, with the splitting reaching even a few GHz (see, e.g., the $m=3$ and $-3$ azimuthal SW eigenmodes in  Fig.~\ref{fig:skyrmion_dynamicsa}(b)). Interestingly, a similar effect was predicted for the vortex state in the ferromagnetic ring with increasing magnitude of the out-of-plane external magnetic field or anisotropy. Here, the origin of the splitting effect was attributed to the topological phase acquired by SWs propagating around the ring, different for the opposite directions of propagation~\cite{Dugaev2005}.

\begin{figure}
  \includegraphics[width=\textwidth]{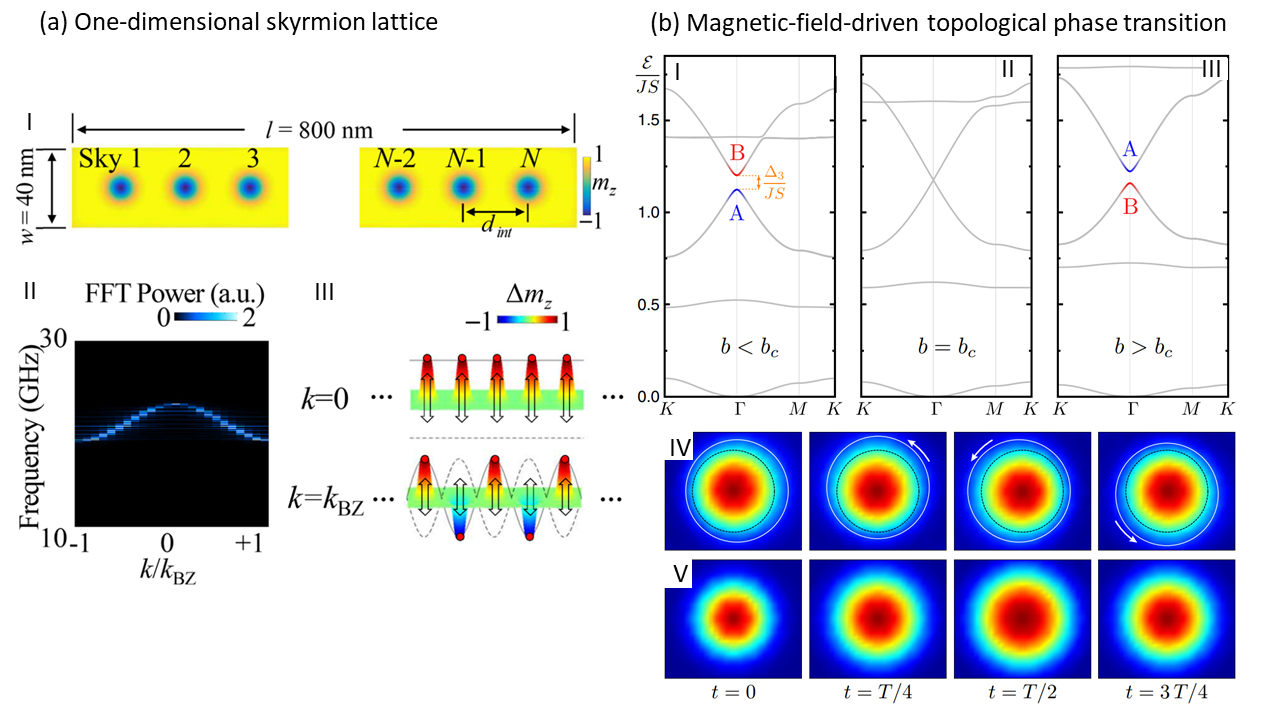}
  \caption{ (a) SWs in 1D skyrmion lattice. (I) The sketch of the 1D skyrmion lattice in a ferromagnetic stripe. Color indicates the out-of-plane component of the magnetization. (II) Dispersion of SWs in 1D skyrmion lattice. (III) Spatial profiles of the $m_{z}$  (out-of-plane) component oscillations in the skyrmion lattice for the collective breathing mode in the Brillouin zone (BZ) center and BZ edge. Figure reproduced with permission from \fullcite{Kim2018CoupledLattices}. Copyright (2022) by the American Institute of Physics. (b) Magnetic-field-driven topological phase transition in skyrmion lattices. (I)–(III) SW spectrum of the 2D SkL at and in the vicinity of the
  topological phase transition for different values of the magnetic field \textit{b} around the critical field value $b_{c}$. The CCW (A) and breathing (B) modes are shown in blue and red, respectively. Time evolution of the CCW (IV) and breathing (V) modes' out-of-plane magnetization. $T$ represents the period of both SW modes. Figure reproduced with permission from~\fullcite{Diaz2020ChiralFields}, under the terms of license CC BY 4.0.}
  \label{fig:skyrmion_dynamicsb}
\end{figure}

The SW dispersion relation in one-dimensional chain of nanodots with skyrmions was investigated theoretically by Mruczkiewicz \textit{et al.} in Ref.~\cite{Mruczkiewicz2016CollectiveCrystal}. The modes forming the collective magnon bands were identified as breathing and clockwise (for assumed skyrmion polarization) gyrotropic SW modes, which correspond to the high- and low-frequency modes, respectively. Later, collective band of the breathing modes in one-dimensional skyrmion chain periodically arranged in a Pt/Co/Ta multilayered thin-film nanostripe (see, Fig.~\ref{fig:skyrmion_dynamicsb} (a)) have been studied by Kim \textit{et al.} in Ref.~\cite{Kim2018CoupledLattices}. They showed that, due to expansion and contraction of the skyrmion cores, the anti-phase breathing oscillations of the neighbouring skyrmions show lower energy than their in-phase oscillations. This means that the band incurs higher energy at the Brillouin zone center ($k=0$) when both cores expand in the same direction than at the Brillouin zone boundary (see, Fig.~\ref{fig:skyrmion_dynamicsb} (a, subplots II-III)). Interestingly, the excited breathing mode propagates with a negative group velocity of relatively high magnitude up to 340 m/s through the 1D lattice. Therefore, collective breathing modes in the skyrmion lattices can serve as information carriers. Importantly, the dispersion relation can be controlled by an external magnetic field that changes the size of the skyrmions or by altering the inter-skyrmion distance. With optimization of these parameters, the group velocity can reach  values even up to 700 m/s. 

The above discussion  of collective excitations of skyrmions undercover the interesting nature and possible utilization of skyrmion dynamics. Due to the fact that the SW band structures and bandgaps of SkLs can be switched dynamically by changing the skyrmion states, the modulation and transition of SW modes from nontrivial to trivial topologies  became a new topic of recent papers~\cite{Wang2020DynamicallyLattice}. For example, very recently, Diaz, \textit{et al.} in Ref.~\cite{Diaz2020ChiralFields}, studied controllable switching of chiral and topologically protected magnonic edge states in a ferromagnetic 2D SkL. They showed that a topological phase transition can occur in the SW spectrum of a N\'eel-type skyrmion lattice just by changing an external magnetic field (see, Fig.~\ref{fig:skyrmion_dynamicsb} (b)). As shown in the sequence of plots in Fig.~\ref{fig:skyrmion_dynamicsb} (b, subplots IV-V), the CCW and breathing modes approach each other at the Brillouin zone center, when the magnetic field decreases. At some specific field value, i.e., at the critical field, the phase transition happens and a low-energy SW gap closes. By further decreasing the field, we observe a reopening of the gap, but importantly, below the critical field,  the bottom band acquires a nonzero Chern number. For the finite system, the two topologically protected magnonic edge states can exist within this gap. The bottom panel of Fig.~\ref{fig:skyrmion_dynamicsb} (b) presents snapshots of the time evolution of the out-of-plane magnetization of the modes from the two considered bands, clearly showing their azimuthal and breathing character. Although the study was performed for 2D systems while neglecting magnetostatic interactions, it reveals interesting property that by applying a magnetic field, a topological phase transition and the corresponding low-energy chiral magnonic edge states can be controlled, promising utilization of robust directional magnon spin currents.

\subsection{Skyrmions motion} \label{motion}

Another appealing feature of the  magnetic skyrmions is the possibility to move them along the track, which can be realized by electric current via spin transfer~\cite{Osca2021TorqueInteraction,Osca2020SkyrmionNanowire,Masell2020Spin-transferCurrents} or spin-orbit torques~\cite{Woo2017Spin-orbitMicroscopy,Zeng2020DynamicsTorque,Montoya2018Spin-orbitTemperature,Nakatani2021DiscriminationOperation}. This feature is particularly interesting from an application point of view because skyrmion's motion is driven by a much lower current density relative to a domain wall's motion~\cite{Nagaosa2013TopologicalSkyrmions,Iwasaki2013UniversalMagnets,Yu2012SkyrmionDensity}. Thus, skyrmions seem to be good candidates for the use  of racetrack magnetic memories~\cite{Fert2017MagneticApplications,Gobel2019OvercomingEffect,Bessarab2018LifetimeSkyrmions,Tomasello2014AMemories,Pan2019ANetwork,Kang2018ASkyrmion}. However, when skyrmions are driven along the straight stripe, they  experience a transverse deflection because of the Magnus force~\cite{Litzius2017SkyrmionMicroscopy,Chen2017SkyrmionApplications,Komineas2015SkyrmionFerromagnets}. This phenomenon, called skyrmion-Hall effect, is regarded as a drawback in-memory applications because eventually, it might end up in destroying the skyrmion at the edge of the track. Intensive research is currently underway to minimize the skyrmion Hall effect, involving different approaches, such as braking skyrmion symmetry~\cite{Zelent2022StabilizationNanostructures,Xia2020DynamicsTorque,Wilson2012ExtendedFilms,Cheng2021EllipticalSpeed}, exploitation of the  synthetic antiferromagnets or utilizing skyrmion transport in pattered structures~\cite{Legrand2020Room-temperatureAntiferromagnets,Tomasello2017PerformanceSkyrmion}.

\begin{figure}
  \includegraphics[width=1.0\textwidth]{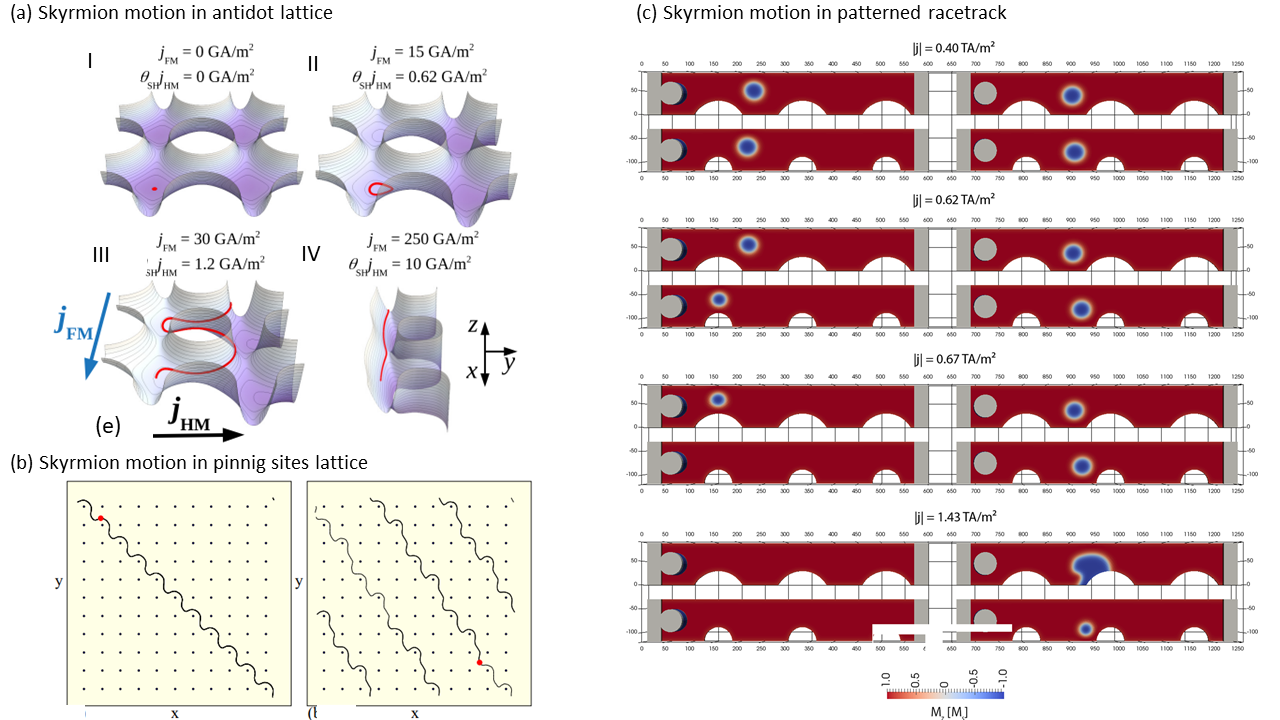}
  \caption{(a) Figure presents effective potential $V_{\textrm{eff}}$ combining the effects of antidot potential $V$ and driving current on the skyrmion motion in the undamped sample. (I)–(IV) Three types of skyrmion trajectories resulting from the applied current. Red curves show the corresponding skyrmion trajectories starting from the potential minimum (valley). Figure reproduced with permission from \fullcite{Feilhauer2020ControlledLattice}. Copyright (2022) by the American Physical Society. (b) Skyrmion trajectories (lines) for N\'eel-type skyrmion (red circle) moving through a periodic array of pinning sites (black dots). Figure reproduced with permission from \fullcite{Reichhardt2015QuantizedSubstrate}. Copyright (2022) by the American Physical Society. (c) Skyrmion movement as a function of applied current density and pinning sites size (in the form of antidot with radius diameter 60 and 120 nm). A positive current density is applied at the right lead (left column) and a negative current is applied at the right lead (right column). Figure reproduced with permission from~\fullcite{Suess2019SpinMemory} under the terms of license CC BY 4.0.}
  \label{fig:skyrmion_movement}
\end{figure}

The above-mentioned skyrmion Hall effect, while undesirable in many cases, can be also practically exploited. Whenever a moving skyrmion meets a pinning site or a barrier (for example an antidot), the Magnus force~\cite{Stone1996MagnusSystems,Fert2017MagneticApplications},  strongly influences the skyrmion motion, causing acceleration effects and skyrmion deflection~\cite{Reichhardt2021StaticsReview,Vizarim2021GuidedInterfaces,Zhang2015Skyrmion-skyrmionMemory,Gonzalez-Gomez2019AnalyticalDefects,Castell-Queralt2020DeterministicRacetracks}. This is a result of the fact that the direction of the applied current and the corresponding Magnus force creates a perpendicular velocity component, while finite damping aligns the skyrmion velocity in the direction of the current~\cite{Reichhardt2021StaticsReview}.  

Feilhauer \textit{et al.} in Ref.~\cite{Feilhauer2020ControlledLattice} showed theoretically that the skyrmions can travel through the channels of ADL. While the electric current has a fixed direction of flow, due to the nontrivial interaction between the repulsive potential introduced by the antidots, the skyrmion Hall effect, and the non-uniform current distribution, a full control of the skyrmion motion can be achieved. The researchers demonstrated that a structured CoFeB layer as an ADL, when magnetized perpendicularly to the film plane can be used to control the skyrmion's motion. In particular, using a rectangular electric current pulse of sufficient density and width, they showed that a skyrmion can be transported between individual valleys of the effective potential with a motion controlled in both the longitudinal and transverse directions. They recognize three types of skyrmion trajectories. Assuming the skyrmion starts from the bottom of the valley presented in Fig.~\ref{fig:skyrmion_movement} (a-I), the skyrmion oscillates inside the valley (a-II), the skyrmion escapes the starting valley, and passes around an antidot to the neighbouring valley (a-III), the skyrmion passes to the neighbouring valley directly through the saddle point between the valleys avoiding the skyrmion motion around the antidot (a-IV). Thus, the skyrmions can be steered to move into almost any position of the ADL by carefully directing the current pulse. The skyrmion can be moved to the desired location with the right sequence of electrical impulses, which is a very important property for applications. 

Similar study on skyrmion movement around pinning sites was performed by Reichhardt \textit{et al.} in Ref.~\cite{Reichhardt2022CrystalsTwist} and Suess \textit{et al.} in Ref.~\cite{Suess2019SpinMemory}. Reichhardt \textit{et al.} discussed semi-analytically the dynamics of a skyrmion moving over a two-dimensional array of pinning sites, utilizing simulations of a particle-based skyrmion model. In particular, they examined the role of the nondissipative Magnus term on the driven motion and the resulting skyrmion velocity-force curves. In Fig.~\ref{fig:skyrmion_movement} (b) for given magnetic and current parameters, the skyrmion follows a sinusoidal trajectory, moving by one saddle point in the $x$-direction and one saddle point in the $y$-direction in a single current period, when the perpendicular (skyrmion Hall effect) and parallel forces (current) acting on the skyrmion are equal. The right hand subplot in Fig.~\ref{fig:skyrmion_movement} (b) illustrates the skyrmion trajectory when the strength of both of these forces are not equal, resulting in different trajectories of the skyrmion. Fig.~\ref{fig:skyrmion_movement} (c) presents the simulated results by Suess \textit{et al.} on skyrmion trajectory for the two different geometries, with a different one-dimensional chain of the pinning sites in dependence on  driving current density and current sign. Their study shows that the skyrmion trajectory in this system strongly depends on the pinning site size (like an anitdot radius), and the current density. In addition, in all presented here papers, they have shown that the movement of a skyrmion can exhibit an acceleration effect, in which the interaction of the skyrmion with a pinning site can speed up the skyrmion in such a way that its velocity is higher than the value that would be induced by an external electric current alone~\cite{Suess2019SpinMemory,Reichhardt2022CrystalsTwist}.

\chapter{Concluding remarks and future challenges}

This chapter reviews selected studies on the magnetization dynamics of patterned ferromagnetic thin films with PMA, which have been a subject of intense research for the past several years. We foresee several possible practical applications of such systems, including the creation of a basis for  reconfigurable magnonic circuits with isotropic SW propagation, and systems for exploiting skyrmions for, i.e., racetrack memories and logic operations~\cite{Gobel2021,Luo2021SkyrmionApplications,Reichhardt2021StaticsReview,Back2020TheRoadmap}. We recognize these two  properties as key characteristics of patterned PMA thin films that are indicated in many studies, but have not yet been fully researched or put into practice.

Theoretical and numerical studies show that SW bands in magnonic and skyrmionic crystals exhibit interesting characteristics such as dynamic tunability or a possibility of existence of topologically nontrivial SW modes~\cite{Wu2021SizeCrystals,Mochizuki2012Spin-waveCrystals,Zang2011DynamicsFilms,Chen2021SkyrmionCrystals,Wang2021StripeCrystals}. Moreover, the thin layers and multilayers of PMA offer promising properties that enable the creation of a system with a programmable, non-uniform magnetization texture, including periodic magnetization patterns~\cite{Mahmoud20204-outputGate,Mahmoud2020IntroductionComputing,Chen2019SpinAnisotropy,Hamalainen2018ControlWall,Goebiewski2022Self-ImagingTables,}. However, the experimental verification of the theoretical predictions has so far been very limited and, for example, includes the lack of experimental confirmation of full magnonic bandgaps in ADL structures, or the formation of band gaps in thin films with periodic stripe domains.

The other interesting area, in which PMA materials with patterned architecture have found interest, is related to magnetic skyrmions. This research area is still very active, and recent works show great progress~\cite{Luo2021SkyrmionApplications,Gobel2021}. In particular, we  discussed the dynamic modes in skyrmion, skyrmion lattices, the proposal for atypical  skyrmion nucleation with the use SWs, and the mobility of skyrmions within patterned nanostructures~\cite{Chen2022SkyrmionDeformation,Komineas2015SkyrmionFerromagnets,Beg2017DynamicsNanostructures,Iwasaki2013Current-inducedGeometries,Nagaosa2013TopologicalSkyrmions,Woo2018Current-drivenFilms}. In these cases, one of the key shortcomings of the experimental studies of dynamic effects is the lack of confirmation of the static magnetic configuration, which is often assumed solely on the basis of micromagnetic simulations. 

The mobility of skyrmions is considered one of their advantages for application in racetrack memories, but their motion is  strongly influenced by Magnus force, local defects, and pinning sites~\cite{Reichhardt2021StaticsReview,Liu2020Current-DrivenDeformation,Tokura2021MagneticMaterials}. We review here the less known use of skyrmion, i.e., skyrmion transport in the periodically patterned films. It has been shown that a properly designed sequence of electric current pulses allows for full control over the skyrmion's movement in a two-dimensional system of pinning centers or antidots. This indicates that these scenarios can be used to design logical operations~\cite{Luo2021SkyrmionApplications,Zhang2015TopologicalCommunication,Zhou2019MagneticConcepts,Chauwin2019SkyrmionComputation}.  

A common challenge in the further development indicated above is the development of experimental methods, for the detection of magnetization textures as well as their dynamics. As an alternative to conventional methods such as ferromagnetic resonance and Brillouin light scattering~\cite{Sebastian2015Micro-focusedNanoscale,Madami2012ApplicationSystems,Banerjee2017MagnonicSimulations}, magnetization textures and skyrmions can be studied using more sophisticated techniques, like time-resolved X-ray imaging~\cite{Cheng2017Phase-resolvedMicroscopy,Gro2022ImagingLattices,Litzius2017SkyrmionMicroscopy,Woo2017Spin-orbitMicroscopy,Nolle2012Note:T,Woo2018DeterministicMicroscopy,Nakajima1998PerpendicularStudy,Grafe2020DirectLens}, which require large-scale facilities. In recent years, work has been carried out on 3D tomography of magnetic textures~\cite{Donnelly2017Three-dimensionalNanotomography,Wolf2021UnveilingTubes,Fischer2017X-rays3D,Donnelly2020Time-resolvedDynamics}, which may also be used for time-resolved measurements in the future.  
 
Thus, the milestone in the advancement of research on the properties of skyrmions is the further development of techniques for detecting magnetic textures and their dynamics, with high spatial and frequency resolution, as well as methods for fabricating structured ferromagnetic materials. The state-of-the-art methods, although capable of observing skyrmions with a high spatial resolution, is limited to low frequencies, thereby limiting the range of SWs possible for observation.

To summarize, there has been mostly theoretical research on spin dynamics in thin patterned films with PMA, while experiments are rare, mainly due to the inhomogeneities in samples, challenges in patterning while keeping the PMA, and other magnetic properties intact or the increase in magnetic damping in materials containing PMA. Although we have found a number of interesting effects of application perspectives for structured samples from PMA, there are still a number of challenges to be solved and questions to be answered. However, it is expected that this research field will see a significant boost in the next few years. The two key drawbacks of these systems are high damping and defects, which make material development and improvement of sample fabrication technology critical for their use in magnonics and spintronics.
\include{concluding_remarks}

\chapter{Acknowledgments}
The research has received funding from National Science Centre of Poland, Grant No.~UMO-2018/30/Q/ST3/00416. AB gratefully acknowledges financial support from the Department of Science and Technology, Govt. of India under Grant No. DST/NM/TUE/QM-3/2019-1C-SNB. The simulations were partially performed at the Poznan Supercomputing and Networking Center (Grant No.~398).

\printbibliography
\end{document}